\documentclass[12pt]{article}
\usepackage{amsmath}
\usepackage{graphicx}
\usepackage{natbib}
\usepackage{url} 

\usepackage{hyperref}

\newcommand{\blind}{0}

\usepackage{amsfonts}
\usepackage{amsthm}

\theoremstyle{plain}
\newtheorem{theorem}{Theorem}[section]
\newtheorem{corollary}[theorem]{Corollary}
\theoremstyle{definition}
\newtheorem{definition}{Definition}
\theoremstyle{remark}
\newtheorem{remark}{Remark}

\begin{document}

\def\spacingset#1{\renewcommand{\baselinestretch}%
{#1}\small\normalsize} \spacingset{1}


\if0\blind
{
  \title{\bf Correcting a Nonparametric Two-sample Graph Hypothesis Test for
    Graphs with Different Numbers of Vertices with Applications to Connectomics}

  \author{Anton A. Alyakin,\thanks{
      Anton A. Alyakin is Assistant Research Engineer
      aalyaki1@jhu.edu), Joshua Agterberg is PhD student (
      jagterb1@jhu.edu), Hayden S. Helm is Adjunct Assistant Research Engineer
      (hhelm2@jhu.edu), and Carey E. Priebe is Professor (
      cep@jhu.edu), Department of Applied Mathematics and Statistics, Johns
      Hopkins University, Baltimore, MD 21218. Carey E. Priebe is the corresponding author. }
    \\
    Joshua Agterberg, \\
    Hayden S. Helm, \\
    and \\
    Carey E. Priebe \\
    Department of Applied Mathematics and Statistics, Johns Hopkins University}
  \maketitle
} \fi

\if1\blind
{
  \bigskip
  \bigskip
  \begin{center}
    {\LARGE\bf Title}
\end{center}
  \medskip
} \fi

\begin{abstract}
  Random graphs are statistical models that have many applications, ranging from
  neuroscience to social network analysis. Of particular interest in some
  applications is the problem of testing two random graphs for equality of
  generating distributions. \cite{nonpar} propose a test for this setting. This
  test consists of embedding the graph into a low-dimensional space via the
  adjacency spectral embedding (ASE) and subsequently using a kernel two-sample
  test based on the maximum mean discrepancy. However, if the two graphs being
  compared have an unequal number of vertices, the test of \cite{nonpar} may not
  be valid. We demonstrate the intuition behind this invalidity and propose a
  correction that makes any subsequent kernel- or distance-based test valid. Our
  method relies on sampling based on the asymptotic distribution for the ASE. We
  call these altered embeddings the {\it corrected adjacency spectral embeddings
    (CASE)}. We also show that CASE remedies the exchangeability problem of the
  original test and demonstrate the validity and consistency of the test that
  uses CASE via a simulation study. Lastly, we apply our proposed test to the
  problem of determining equivalence of generating distributions in human
  connectomes extracted from diffusion magnetic resonance imaging (dMRI) at
  different scales.
\end{abstract}

\noindent%
{\it Keywords:} Adjacency Spectral Embedding, Latent Position Graph, Random Dot
Product Graph
\vfill

\newpage
\spacingset{1.5} 

\section{Introduction}
\label{sec:intro}
Modeling data as random graphs is ubiquitous in many application domains of
statistics. For example, in neuroscience, it is common to view a connectome as a
graph with vertices representing neurons, and edges representing synapses
\citep{neuroscience-application-1}. In document analysis, the corpus of text can
be viewed as a graph with vertices taken to be documents or authors, and edges
as the citations \citep{document-application-1}. In social network analysis, a
network can be modeled as a graph with vertices being individual actors or
organizations, and edges being representing the degree of communication between
them \citep{social-application-1}.

The first random graph model was proposed in 1959 by E. N. Gilbert. In his short
paper, he considered a graph in which the probability of an edge between any two
vertices was a Bernoulli random variable with common probability $p$
\citep{er-graphs-1-gilbert}. Almost concurrently, Erd\"{o}s and R\'{e}nyi
developed a similar random graph model with a constrained number of edges that
are randomly allocated in a graph. They also provided a detailed analysis of the
probabilities of the emergence of certain types of subgraphs within graphs
developed both by them and Gilbert \citep{er-graphs-3}. Nowadays, the graphs in
which edges arise independently and with common probability $p$ are known as
Erd\"{o}s-R\'{e}nyi (ER) graphs.

Latent position random graph models consitute a diverse class of random graph
models that are much more flexible than the ER model. A vertex in a latent
position graph is associated with an element in a latent space $\mathcal{X}$,
and the probability of an edge between any two vertices is given by a link
function $g : \mathcal{X} \times \mathcal{X} \to [0, 1]$
\citep{latent-graphs-intro}. The model draws inspiration from social network
analysis, in which the members are thought of as vertices, and the latent
positions are differing ``interests''. Latent position random graphs are a
submodel of the independent edge graphs, that is, graphs in which the edge
probabilities are indpendent, conditioned on a matrix of probabilities. The
theory of latent positions graphs is also closely related to that of graphons
\citep{graphons-1}; for discussion on this relationship, see, for example,
\cite{graphons-2} or \cite{graphons-3}.

One example of latent position graphs relevant to this discussion is the random
dot product graph (RDPG). An RDPG is a latent position graph in which the latent
space is an appropriately constrained Euclidian space $\mathbb{R}^d$, and the
link function is the inner product of the $d$-dimensional latent positions
\citep{rdpg-survey}. Despite their relative simplicity, suitably
high-dimensional RDPGs can provide useful approximations of general latent
position and independent edge graphs, as long as their matrix of probabilities
is positive semidefinite \citep{rdpg-approximation}.

The well-known stochastic blockmodel (SBM), in which each vertex belongs to one
of $K$ blocks, with connection probabilities determined solely by block
membership \citep{sbm}, can be represented as an RDPG for which all vertices in
a given block have the same latent positions. Furthermore, common extensions of
SBMs, namely degree-corrected SBMs \citep{dcsbm}, mixed membership SBMs
\citep{mmsbm}, and degree-corrected mixed membership SBMs \citep{dcmmsbm-1} can
also be framed as RDPG. There is, however, a caveat, similar to the one for
approximating independent edge graphs with RDPG: only SBM graphs with positive
semidefinite block probability matrix can be formulated in the context of RDPG.
\cite{grdpg} present a generalization of RDPGs, called the generalized random
dot product graph (GRDPG) that allows to drop the positive semidefiniteness
requirements in both cases. Although the generalization of many estimation and
inference procedures from RDPGs to GRDPGs is straightforward, their theory,
particularly of latent distribution testing, is not yet as developed as that of
RDPG. Thus, we limit the scope of this work to RDPG.

The problem of whether the two graphs are ``similar'' in some appropriate sense
arises naturally in many fields. For example, two different brain graphs may be
tested for the similarity of the connectivity structure \citep{varjavand}, or
user behavior may be compared between different social media platforms. Testing
for similarity also has applications in more intricate network analysis
techniques, such as hierarchical community detection \citep{hsbm, hierarchical}.
Despite the multitude of applications, network comparison is a relatively
nascent field, and comparatively few techniques currently exist \citep{hsbm}.
There have been several tests assuming the random graphs have the same set of
nodes, such as \cite{semipar, omni, graph-comparison-same-1,
  graph-comparison-same-2, graph-comparison-same-3}, and
\cite{graph-comparison-same-4}. Other approaches designed for fixed models and
related problems, include \cite{graph-comparison-other-1,
  graph-comparison-other-2, graph-comparison-other-3, graph-comparison-other-4,
  graph-comparison-other-5, graph-comparison-other-6, graph-comparison-other-7}
and \cite{dcmmsbm-2}, to name a few. In \cite{graph-comparison-other-8}, the
authors formulate the two-sample testing problem for graphs of different orders
more generally.

Of particular interest is \cite{nonpar}, in which the authors propose a
nonparametric test for the equality of the generating distributions for a pair
of random dot product graphs. This test does not require the graphs to have the
same set of nodes or be of the same order. It relies on embedding the adjacency
matrices of the graphs into Euclidean space, followed by a kernel two-sample
test of \cite{gretton-survey} performed on these embeddings. The exact
finite-sample distribution of the test statistics is unknown, but it can be
estimated using a permutation test, or approximated using the
$\chi^2$-distribution. Unfortunately, despite the theorem stating that in the
limit, even for graphs of differing orders, the statistic using the two
embeddings converges to the statistic obtained using the true but unknown latent
positions, the test is not always valid for finite graphs of differing orders.

The invalidity arises from the fact that the approximate finite-sample variance
of the adjacency spectral embedding depends on the number of vertices
\citep{avanti-clt}. Hence, the distributions of the estimates of the latent
positions for the two graphs might not be the same, even if the true
distributions of the latent positions are equivalent. The test of
\cite{gretton-survey} is sensitive to the differences induced by this
incongruity and as a result may reject more often than the intended significance
level. In this work, we present a method for modifying the embeddings before
computing of the test statistic. Using this correction makes the test for the
equivalence of latent distributions valid even when the two graphs have an
unequal number of vertices.

The remainder of the paper is structured as follows. In Section
\ref{sec:preliminaries}, we review the random dot product graph, and discuss its
relationship with Erd\"{o}s-R\'{e}nyi, stochastic blockmodel and other random
graph models. We also discuss results associated with the adjacency spectral
embedding of an RDPG, such as consistency for the true latent positions and
asymptotic normality, and we review the original nonparametric two-sample
hypothesis test for the equality of the latent distributions. There we also
briefly discuss generalizing the ASE procedure to weighted and/or directed
graphs. Then, in Section \ref{sec:fix}, we give an intuition as to why this test
increases in size as the orders of the two graphs diverge from each other. We
also present our approach to correcting the adjacency spectral embeddings in a
way that makes them exchangeable under the null hypothesis of the test for the
equivalence of the latent distribution. We demonstrate the validity and
consistency of the test that uses the corrected adjacency spectral embeddings
across a variety of settings in Section \ref{sec:experiments}. In Section
\ref{sec:real_data}, we demonstrate that failing to correct for the difference
in distributions can lead to significant inferential consequences in real world
applications, such as the setting of brain graphs obtained from diffusion
magnetic resonance imaging (dMRI). Furthermore, we show that the test is able to
meaningfully differentiate between scans within the same subject and different
subjects. We conclude and discuss our findings in Section \ref{sec:conclusion}.

\subsection{Notation}
We use the terminology ``order'' for the number of vertices in a graph. We
denote scalars by lowercase letters, vectors by bold lowercase letters and
matrices by bold capital letters. For example, $c$ is a scalar, $\boldsymbol{x}$
is a vector, and $\boldsymbol{H}$ is a matrix. For any matrix $\boldsymbol{H}$,
we let $\boldsymbol{H}_{ij}$ denote its $i, j$th entry. For ease of notation, we
also denote $\boldsymbol{H}_i$ to be the column vector obtained by transposing
the $i$-th row of $\boldsymbol{H}$. Formally, $\boldsymbol{H}_i =
\left(\boldsymbol{H}_{i \cdot}\right)^T$. In the case where we need to consider
a sequence of matrices, we will denote such a sequence by
$\boldsymbol{H}^{(n)}$, where $n$ is the index of the sequence. Whether a
particular scalar, vector or a matrix is a constant or a random variable will be
stated explicitly or be apparent from context. Unbold capital letters denote
sets or probability distributions. For example, $F$ is a probability
distribution. The exception to this rule is $K$ which is always used to denote
the number of blocks in a stochastic blockmodel.

\section{Preliminaries}
\label{sec:preliminaries}
\subsection{Models}
We begin by defining random dot product graphs.
\begin{definition}[$d$-dimensional random dot product graph (RDPG)]
  Let $F$ be a distribution on a set $\mathcal{X} \subset \mathbb{R}^d$ such
  that $\langle \boldsymbol{x}, \boldsymbol{x}' \rangle \in [0, 1]$ for all
  $\boldsymbol{x}, \boldsymbol{x}' \in \mathcal{X}$. We say that
  $(\boldsymbol{X}, \boldsymbol{A}) \sim RDPG(F, n)$ is an instance of a random
  dot product graph (RDPG) if $\boldsymbol{X} = [\boldsymbol{X}_1, \hdots ,
  \boldsymbol{X}_n]^T$ with $\boldsymbol{X}_1. \hdots, \boldsymbol{X}_n
  \stackrel{iid}{\sim} F$ and $\boldsymbol{A} \in \left\{ 0, 1 \right\}^{n
    \times n}$ is a symmetric hollow matrix whose entries in the upper triangle
  are conditionally independent given $\boldsymbol{X}$ and satisfy
  \begin{align*}
    \boldsymbol{A}_{ij} \vert \boldsymbol{X} \sim Bernoulli(\boldsymbol{X}_i^T \boldsymbol{X}_j) && i < j.
  \end{align*}
  We refer to $\boldsymbol{X}_1. \hdots, \boldsymbol{X}_n$ as the {\it latent
    positions} of the corresponding vertices.
\end{definition}

\begin{remark}
It is easy to see that if  $(\boldsymbol{X}, \boldsymbol{A}) \sim RDPG(F, n)$,
then $ E [\boldsymbol{A} \vert \boldsymbol{X}] = \boldsymbol{X} \boldsymbol{X}^T $.
\end{remark}

\begin{remark}
  Nonidentifiability is an intrinsic property of random dot product graphs. For
  any matrix $\boldsymbol{X}$ and any orthogonal matrix $\boldsymbol{W}$, the
  inner product between any rows $i, j$ of $\boldsymbol{X}$ is identical to that
  between the rows $i, j$ of $\boldsymbol{X}\boldsymbol{W}$. Hence, for any
  probability distribution $F$ on $\mathcal{X}$ and orthogonal operator
  $\boldsymbol{W}$, the adjacency matrices $\boldsymbol{A}$ and
  $\boldsymbol{B}$, generated according to $(\boldsymbol{X}, \boldsymbol{A})
  \sim RDPG(F, n)$ and $(\boldsymbol{Y}, \boldsymbol{B}) \sim RDPG(F \circ
  \boldsymbol{W}, n)$, respectively, are identically distributed.
  Here, the notation $F \circ \boldsymbol{W}$ means that if $\boldsymbol{X}
  \sim F$, then $\boldsymbol{X} \boldsymbol{W} \sim F \circ \boldsymbol{W}$.
\end{remark}

Constraining all latent positions to a single value leads to an
Erd\"{o}s-R\'{e}nyi (ER) random graph.
\begin{definition}[Erd\"{o}s-R\'{e}nyi graphs (ER)]
  We say that a graph $(\boldsymbol{X}, \boldsymbol{A}) \sim RDPG(F, n)$ is an
  Erd\"{o}s-R\'{e}nyi (ER) graph with an edge probability $p^2$ if $F$ is a
  pointmass at $p$. In this case, we write $\boldsymbol{A} \sim ER (n, p^2)$.
\end{definition}

Another random graph model that can be framed in the context of random dot
product graphs is the stochastic blockmodel (SBM) \citep{sbm}. In the SBM, the
vertex set is thought of as being partitioned into $K$ groups, called blocks,
and the probability of an edge between two vertices is determined by their block
memberships. The partitioning, or assignment, of the vertices is usually itself
random and mutually independent. Formally, we can define SBMs in terms of the
RDPG model as follows.
\begin{definition}[(Positive semidefinite) stochastic blockmodel (SBM)]
  \label{sbm_definition}
  Denote $\delta_{\boldsymbol{z}}$ as the Dirac delta measure at
  $\boldsymbol{z}$. We say that a graph $(\boldsymbol{X}, \boldsymbol{A}) \sim
  RDPG(F, n)$ is a {\it (positive semidefinite) stochastic blockmodel (SBM)}
  with $K$ blocks if the distribution $F$ is a mixture of $K$ point masses,
  \begin{align}
    F = \sum_{i=1}^{K} \pi_i \delta_{\boldsymbol{Z}_i},
  \end{align}
  where $\boldsymbol{\pi} = \left[\pi_1, \hdots, \pi_K \right] \in (0, 1)^K$
  satisfying $\sum_{i=1}^K \pi_i = 1$, and the distinct latent positions are
  given by $\boldsymbol{Z} = \left[ \boldsymbol{Z}_1, \hdots, \boldsymbol{Z}_K
  \right]^T \in \mathbb{R}^{K \times d}$, with $\boldsymbol{Z}_i^T
  \boldsymbol{Z}_j \in [0, 1]$ $\forall i, j$. In this case we also write
  $\boldsymbol{A} \sim SBM \left(n, \boldsymbol{\pi}, \boldsymbol{P} \right)$,
  where $\boldsymbol{P} := \boldsymbol{Z} \boldsymbol{Z}^T \in \mathbb{R}^{K
    \times K}$. The matrix $\boldsymbol{P}$ is often referred to as {\it block
    probability matrix}.
\end{definition}

\begin{remark}
  We note that almost everywhere below we use the terms SBM and positive
  semidefinite SBM interchangeably, as only positive semidefinite block
  probability matrices can be represented as a product of a matrix of latent
  positions with transpose of itself, and thus only they can be defined in terms
  of the RDPG model. We emphasize, however, that the work of \cite{grdpg} on the
  generalized random dot product (GRDPG) extends the construction of RDPG via
  the indefinite inner product to encompass indefinite SBM and the
  generalizations thereof.
\end{remark}

There are two common generalizations of the stochastic blockmodel:
degree-corrected stochastic blockmodel \citep{dcsbm} and mixed-membership
stochastic blockmodel \citep{mmsbm}. We present these two models below. The
presentations are different from the ones many readers may be familiar with
because we present them under the RDPG framework. These definitions coincide
with the ones in literature, as covered in \cite{ase-consistency-3, mmsbm-rdpg,
  grdpg}.

\begin{figure}[t]
  \begin{center}
    \includegraphics[width=0.95\textwidth]{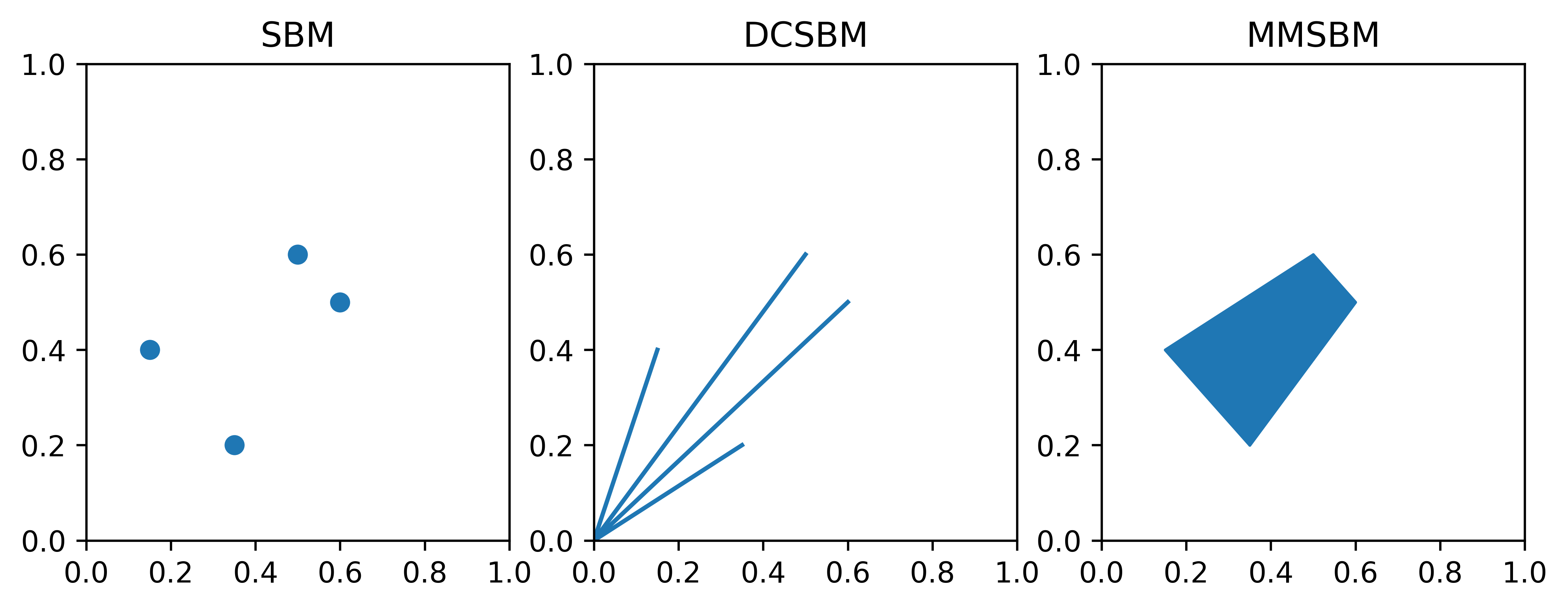}
  \end{center}
  \caption{Visualization of the valid latent positions of an arbitrary
    2-dimensional SBM with $K=4$ (left), valid latent positions of a DCSBM with
    the same $\boldsymbol{Z}$ (center) and valid latent positions of an MMSBM
    with the same $\boldsymbol{Z}$ (right). All three are examples of RDPGs.}
  \label{fig:sbm_types}
\end{figure}

The degree-corrected stochastic blockmodel allows for vertices within each
block to have different expected degrees, which makes it more flexible
than the standard SBM and a popular choice to model network data \citep{dcsbm,
  ase-consistency-3}.

\begin{definition}{(Degree-corrected SBM)}
  We say that a graph $(\boldsymbol{X}, \boldsymbol{A}) \sim RDPG(F, n)$ is a
  degree-corrected SBM (DCSBM) with $K$ blocks, if there exists a distribution
  $F_m$, which is a mixture of $K$ point-masses $\boldsymbol{Z}_1, \hdots,
  \boldsymbol{Z}_K$, as in Definition \ref{sbm_definition}, and a distribution
  $F_c$ on $[0, 1]$, such that for all $\boldsymbol{X}_i$, there exists
  $\boldsymbol{Y}_i \sim F_m$ and $c_i \sim F_c$, such that $\boldsymbol{X}_i =
  c_i \boldsymbol{Y}_i$.
\end{definition}

That is, any latent position of a vertex in a DCSBM graph can be decomposed into
a point $\boldsymbol{Y}_i$, chosen among one of the $K$ shared points
$\boldsymbol{Z}_1, \hdots, \boldsymbol{Z}_K$, and a scalar $c_i$. Note that
there is no requirement on $\boldsymbol{Y}_i$ and $c_i$ to be independent from
each other. In other words, the distributions on degree corrections can depend
on the block assignments. In essence, the DCSBM generalizes the SBM from an RDPG
with a distribution of latent positions over a finite number of points to an
RDPG with a distribution of latent positions over a finite number of rays from
the origin. Of course, not every point on these rays needs to be in the support
of this distribution. Restraining $F_c$ to a point-mass at unity recovers the
regular SBM. See left and central panels of Figure \ref{fig:sbm_types} for a
visualization comparing the latent distributions of SBM and DCSBM.

On the other hand, the mixed membership SBM offers more flexibility in block
memberships by allowing each vertex to be in a mixture of blocks \citep{mmsbm}.
\begin{definition}{(Mixed-membership SBM)}
  Denote $\Delta^{d \times 1}$ to be the space of the $(d+1)$-dimensional column
  vectors starting at the origin and terminating in the $d$-dimensional unit
  simplex. We say that a graph $(\boldsymbol{X}, \boldsymbol{A}) \sim RDPG(F,
  n)$ is a mixed-membership SBM (MMSBM) with $K$ blocks, if there exists a
  matrix $\boldsymbol{Z} = \left[ \boldsymbol{Z}_1, \hdots, \boldsymbol{Z}_K
  \right]^T \in \mathbb{R}^{K \times d}$ and a distribution over $\Delta^{(K-1)
    \times 1}$, denoted as $F_m$, such that for each $\boldsymbol{X_i}$, there
  exists $\boldsymbol{v}_i \sim F_m$ and $\boldsymbol{X_i} = \boldsymbol{Z}^T
  \boldsymbol{v}_i$.
\end{definition}

That is, any latent position of a vertex in an MMSBM is a convex combination of
$K$ shared points, $\boldsymbol{Z}_1, \hdots, \boldsymbol{Z}_K$. The MMSBM
generalizes the SBM from an RDPG with latent positions coming from a
finite-dimensional mixture of point-masses to an RDPG with latent positions
having a distribution over a convex hull formed by a finite number of points.
See left and right panels of Figure \ref{fig:sbm_types} for a visualization of
thereof. Once again, the whole convex hull needs not be in the support of this
distribution. If one constrains $F_m$ to only have support on a finite set of
vectors with 1 in a single entry and 0 in all other, $F_m$ collapses to a
distribution of point-masses and the model agrees exactly with SBM.

\begin{remark}
\label{rpdg_sbm_equivalence}
For graphs with one-dimensional latent positions, any RDPG model is both a DCSBM
with a single block and an MMSBM with two blocks. To see this, note that the
latent positions all take values in $[0, 1]$ (or equivalently $[-1, 0]$). This
region can be thought of as either a single line segment starting from the
origin or as a one-dimensional convex hull between $0$ and $1$.
\end{remark}
\begin{remark}
  \cite{dcmmsbm-1} introduced a model that has both the degree heterogeneity of
  the DCSBM and the flexible memberships of MMSBM. This model can also be
  formulated in terms of the RDPG. See, for example, Definition 6 of
  \cite{on-two-sources}.
\end{remark}

We reiterate that the SBM with $K$ blocks is a submodel of both the $K$-block
DCSBM and the $K$-block MMSBM. Furthermore, both the $K$-block DCSBM and the
$K$-block MMSBM are submodels
of 
an RDPG with latent positions in at most $K$ dimensions. Hence, any test for the
equality of the latent distributions that is consistent in the RDPG setting will
be able to meaningfully distinguish between two graphs generated from two
different model subspaces, or between graphs from the same model subspace but
with different parameters; for example, between a MMSBM and an SBM, or between
two SBMs with different block-probability matrices.

\subsection{Adjacency spectral embedding}
Inference on random dot product graphs relies on having good estimates of the
latent positions of the vertices. One way to estimate the latent positions is to
use the adjacency spectral embedding of the graph, defined as follows.
\begin{definition}[Adjacency spectral embedding (ASE)]
  Let $\boldsymbol{A}$ have eigendecomposition
  \begin{align*}
    \boldsymbol{A}
    = \boldsymbol{U} \boldsymbol{\Lambda} \boldsymbol{U}^{\top}
    + \boldsymbol{U}_{\perp} \boldsymbol{\Lambda}_{\perp} \boldsymbol{U}_{\perp}^T
  \end{align*}
  where $\boldsymbol{U}$ and $\boldsymbol{\Lambda}$ consist of the top $d$
  eigenvectors and eigenvalues (arranged by decreasing magnitude) respectively,
  and $\boldsymbol{U}_{\perp}$ and $\boldsymbol{\Lambda}_{\perp}$ consist of the
  bottom $n-d$ eigenvectors and eigenvalues respectively. The adjacency
  spectral embedding of $\boldsymbol{A}$ into $\mathbb{R}^d$ is the $n \times d$
  matrix
  \begin{align*}
      \hat{\boldsymbol{X}} := \boldsymbol{U} \vert \boldsymbol{\Lambda} \vert^{1/2},
  \end{align*}
  where the operator $\vert \cdot \vert$ takes the entrywise absolute value.
\end{definition}

It has been proven in \cite{ase-consistency-1, ase-consistency-2} and
\cite{ase-consistency-3} that the adjacency spectral embedding provides a
consistent estimate of the true latent positions in random dot product graphs.
The key to this result is tight concentrations, in both Frobenius and $2 \to
\infty$ norms, of the ASE about the true latent positions.

\cite{avanti-clt} show that for a $d$-dimensional RDPG with i.i.d. latent
positions, the ASE is not only consistent, but also asymptotically normal, in
the sense that there exists a sequence of $d \times d$ real orthogonal matrices
$\boldsymbol{W}^{(n)}$ such that for any row index $i$, $\sqrt{n}
\left(\boldsymbol{W}^{(n)} \hat{\boldsymbol{X}}_i^{(n)} - \boldsymbol{X}_i^{(n)}
\right)$ converges to a (possibly infinite) mixture of multivariate normals.
\begin{theorem}[RDPG Central Limit Theorem]
  \label{rdpg_clt}
  Let $(\boldsymbol{X}^{(n)}, \boldsymbol{A}^{(n)}) \sim RDPG(F, n)$ be a
  sequence of latent positions and associated adjacency matrices of a
  $d$-dimensional RDPG according to a distribtuion $F$ in an appropriately
  constrained region of $\mathbb{R}^d$. Also let $\hat{\boldsymbol{X}}^{(n)}$ be
  the adjacency spectral embedding of $\boldsymbol{A}^{(n)}$ into
  $\mathbb{R}^d$. Let $\Phi(\boldsymbol{z}, \boldsymbol{\Sigma})$ denote the
  cumulative distribution function for the multivariate normal, with mean zero
  and covariance matrix $\boldsymbol{\Sigma}$, evaluated at $\boldsymbol{z}$.
  Then there exists a sequence of orthogonal $d \times d$ matrices
  $\left(\boldsymbol{W}^{(n)}\right)_{n=1}^{\infty}$ such that for each
  component $i$ and any $\boldsymbol{z} \in \mathbb{R}^d$,
  \begin{align*}
    \lim_{n \to \infty} \mathbb{P}
    \left[n^{1/2} \left(\hat{\boldsymbol{X}}^{(n)} \boldsymbol{W}^{(n)}
    - \boldsymbol{X}^{(n)} \right)_i \leq \boldsymbol{z} \right]
    = \int_{\mathrm{supp} F} \Phi \left(\boldsymbol{z}, \boldsymbol{\Sigma} ( \boldsymbol{x} ) \right) d F (\boldsymbol{x}),
  \end{align*}
  where
  \begin{align*}
    \boldsymbol{\Sigma} (\boldsymbol{x})
    &= \Delta^{-1} \mathbb{E} \left[ (\boldsymbol{x}^T \boldsymbol{X}_1 -
      (\boldsymbol{x}^T \boldsymbol{X}_1)^2) \boldsymbol{X}_1 \boldsymbol{X}_1^T
      \right] \Delta^{-1}
  \end{align*}
  and $\Delta = \mathbb{E} \left[\boldsymbol{X}_1 \boldsymbol{X}_1^T \right]$ is
  the second moment matrix.
\end{theorem}
An intuitive way to restate this result is by identifying that each row
$\hat{\boldsymbol{X}}_i$ of the ASEs $\hat{\boldsymbol{X}}$ is approximately
normal around the true but unknown realization of the latent position of the
vertex:
\begin{align*}
  \hat{\boldsymbol{X}}_i \vert \boldsymbol{X}_i \overset{approx}{\sim} \mathcal{N}
  \left(\boldsymbol{X}_i \boldsymbol{W},
  \frac{\boldsymbol{\Sigma}(\boldsymbol{X}_i)}{n} \right)
\end{align*}
where $\boldsymbol{W}$ is an orthogonal matrix present due to the inherent
orthogonal nonidentifiability of the RDPG.

In our work, we will need to estimate the covariance matrix
$\boldsymbol{\Sigma}(\boldsymbol{X}_i)$. The plug-in principle
\citep{Bickel+Doksum} states that one acceptable method of estimating
$\boldsymbol{\Sigma}(\boldsymbol{X}_i)$ is to use the analogous empirical
moments:
\begin{align}
  \hat{\boldsymbol{\Sigma}} (\hat{\boldsymbol{X}}_i)
  &= \hat{\Delta}^{-1}
    \left(\frac{1}{n} \sum_{j=1}^n \left(
    (\hat{\boldsymbol{X}}_i^T \hat{\boldsymbol{X}}_j
    - (\hat{\boldsymbol{X}}_i^T \hat{\boldsymbol{X}}_j)^2)
    \hat{\boldsymbol{X}}_j \hat{\boldsymbol{X}}_j^T \right) \right)
    \hat{\Delta}^{-1} \label{plug-in-estimator},
\end{align}
where
\begin{align*}
  \hat{\Delta}
  &= \frac{1}{n} \sum_{j=1}^n \hat{\boldsymbol{X}}_j \hat{\boldsymbol{X}}_j^T.
\end{align*}
When we are presented with two or more RDPGs that have the same distribution for
their latent positions, either by assumption or by prior knowledge, we can
leverage this fact and calculate the moments over all graphs at the same time.
Conceptually this is similar to using pooled variance in classical
one-dimensional two-sample inference.

A corollary of the previous result arises when $(\boldsymbol{X}, \boldsymbol{A})
\sim RDPG(F, n)$ is a $K$-block stochastic blockmodel. Then, we can condition on
the event that $\boldsymbol{X}_i$ is assigned to a block $k \in \{1, 2, \hdots,
K \}$ to show that the conditional distribution of $\hat{\boldsymbol{X}}^{(n)}
\boldsymbol{W}^{(n)} - \boldsymbol{X}^{(n)}$ converges to a multivariate normal.
\begin{corollary}
\label{sbm_clt}
Assume the setting and notation of Theorem \ref{rdpg_clt}. Further, assume that
$(\boldsymbol{X}, \boldsymbol{A}) \sim RDPG(F, n)$ is a positive definite
stochastic blockmodel, that is, $F$ is a mixture of $K$ point masses
$\boldsymbol{Z}_1, \hdots, \boldsymbol{Z}_K$, as per Definition
\ref{sbm_definition}. Then there exists a sequence of orthogonal matrices
$\boldsymbol{W}_n$ such that for all $\boldsymbol{z} \in \mathbb{R}^d$ and for
any fixed index $i$,
  \begin{align*}
    \lim_{n \to \infty} \mathbb{P}
    \left[n^{1/2} \left(\hat{\boldsymbol{X}}^{(n)} \boldsymbol{W}^{(n)}
    - \boldsymbol{X}^{(n)} \right)_i \leq \boldsymbol{z}
    \vert \boldsymbol{X}_i = \boldsymbol{Z}_k\right]
    = \Phi \left(\boldsymbol{z}, \boldsymbol{\Sigma} ( \boldsymbol{Z}_k ) \right)
  \end{align*}
\end{corollary}
Consequently, the unconditional limiting distribution in this
setting is a mixture of $K$ multivariate normals \citep{avanti-clt}.

\begin{remark}
  As a special case of Corollary \ref{sbm_clt}, we note that if
  $\boldsymbol{A} \sim ER(n, p^2)$, then the adjacency embedding of
  $\boldsymbol{A}$, $\hat{\boldsymbol{X}}$, satisfies
\begin{align*}
    n^{1/2} (\hat{\boldsymbol{X}}_i - p ) \to \mathcal{N} \left(0, 1 - p^2\right).
\end{align*}
\end{remark}

\subsection{The directed, the weighted, and the unknown dimension ASE}
\label{sec:generalizations}
Although the theory of the ASE and the nonparametric test is predominantly
developed of the setting of undirected unweighted graphs with an assumed known
distribution of the true latent dimension, the real world datasets often require
us to relax those assumptions. This will be the case in our Section
\ref{sec:real_data} which will present an illustrative example using the dMRI
dataset. Graphs in this dataset are weighted and directed, and have an unknown
true distribution of the latent dimension which requires having a modification
to the procedure and interpretation described previously. These modifications to
the statistical procedures involving ASE of the RDPGs that are not unweighted
or/nor undirected are described in more details in Section 6.3 of
\cite{rdpg-survey} where authors apply a clustering algorithm to a dataset of
the larval {\it drosophila} mushroom body connectome which is a directed graph
on four neuron types.

The presence of weights changes the interpretation of the embeddings, as the
inner product no longer represents a probability of an edge, but does not
require modifications to any of the algorithmic procedures. We do, however, need
to define a special adjacency spectral embedding of a directed graph, as the
adjacency matrix is no longer symmetric and thus does not have an
eigendecomposition.

\begin{definition}[Adjacency spectral embedding (ASE) of a directed graph]
  Let $d \geq 1$ and let $\boldsymbol{A}$ be an adjacency matrix of a directed
  graph with $n$ vertices. Let $\boldsymbol{A}$ have singular value
  decomposition
  \begin{align*}
    \boldsymbol{A}
    = \boldsymbol{U} \boldsymbol{\Lambda} \boldsymbol{V}^{\top}
    + \boldsymbol{U}_{\perp} \boldsymbol{\Lambda}_{\perp} \boldsymbol{V}_{\perp}^T
  \end{align*}
  $\boldsymbol{\Lambda}$ is a $d \times d$ diagonal matrix consisting of $d$
  largest singular values and $\boldsymbol{U}$ and $\boldsymbol{V}$ are the
  associated matrices of left andright singular vectors. The adjacency spectral
  embeddingof a directed graph $\boldsymbol{A}$ into $\mathbb{R}^{2d}$ is the
  $n \times 2d$ matrix
  \begin{align*}
    \hat{\boldsymbol{X}} := \left[ \boldsymbol{U} \boldsymbol{\Lambda} ^{1/2}
                             \vert \boldsymbol{V} \boldsymbol{\Lambda}^{1/2} \right].
  \end{align*}
\end{definition}

The scaled left-singular vectors $\boldsymbol{U} \boldsymbol{\Lambda} ^{1/2}$
can be thought of as the ``out-vector'' representation of the directed graph,
and similarly, $\boldsymbol{V} \boldsymbol{\Lambda} ^{1/2}$ can be interpreted
as the ``in-vectors'' \citep{rdpg-survey}. The subsequent inference generally
does not differ in any way after obtaining the ASE of the directed graph.

The ``optimal'' dimension $d$ (or $2d$ in a directed case) to embed into is
often unknown and must be estimated. In general, identifying the ``best'' method
is impossible, as the bias-variance tradeoff demonstrates that, for small $n$,
subsequent inference may be optimized by choosing a dimension smaller than the
true signal dimension, see \cite{impossible} for a clear and concise
illustration of this phenomenon. For a brief discussion of methods applicable to
this problem in the graph embedding setting, see Section 6.3 of
\cite{rdpg-survey}. In our work, we elect to use the automated profile
likelihood-based single value thresholding method of \cite{profile-likelihood}
when the true dimension is unknown (i.e. Section \ref{sec:real_data}). In the
cases when the optimal dimensions of the two graphs being compared are not
equal, we pick the larger of the two. For our simulation study in section
\ref{sec:experiments} we assume that the true dimension is known \textit{a
  priori}.

\subsection{Nonparametric latent distribution test}
\cite{nonpar} present the convergence result of the test statistic in the test
for the equivalence of the latent distributions of two RDPG. One of their main
theorems is presented below.
\begin{theorem}
  \label{thorem:nonpar}
  Let $(\boldsymbol{X}, \boldsymbol{A}) \sim RDPG(F, n)$ and $(\boldsymbol{Y},
  \boldsymbol{B} ) \sim RDPG(G, m)$ be $d$-dimensional random dot product
  graphs. Assume that the distributions of latent positions $F$ and $G$ are such
  that the second moment matrices $\mathbb{E}[\boldsymbol{X}_1
  \boldsymbol{X}_1^T]$ and $\mathbb{E}[\boldsymbol{Y}_1 \boldsymbol{Y}_1^T]$
  each have $d$ distinct nonzero eigenvalues. Consider the hypothesis test
  \begin{align*}
    &H_0 : F = G \circ \boldsymbol{W}
    && \text{for some orthogonal operator } \boldsymbol{W} \\
       &H_A : F \neq G \circ \boldsymbol{W}
    && \text{for all orthogonal operators } \boldsymbol{W}.
  \end{align*}

  Denote by $\hat{\boldsymbol{X}} = \left\{\hat{\boldsymbol{X}}_1, \hdots ,
    \hat{\boldsymbol{X}}_n \right\}$ and $\hat{\boldsymbol{Y}} =
  \left\{\hat{\boldsymbol{Y}}_1,\hdots, \hat{\boldsymbol{Y}}_m \right\}$ the
  adjacency spectral embeddings of $\boldsymbol{A}$ and $\boldsymbol{B}$
  respectively. Recall that a radial basis kernel $\kappa(\cdot, \cdot)$ is any
  kernel such that $\kappa(\boldsymbol{W} \boldsymbol{x}, \boldsymbol{W}
  \boldsymbol{y} ) = \kappa (\boldsymbol{x}, \boldsymbol{y})$ for all
  $\boldsymbol{x}, \boldsymbol{y}$ and orthogonal transformations
  $\boldsymbol{W}$. Define the test statistic
  \begin{align*}
    T_{n,m} \left(\hat{\boldsymbol{X}}, \hat{\boldsymbol{Y}} \right)
    =& \frac{1}{n(n-1)} \sum_{j\neq i} \kappa \left(\hat{\boldsymbol{X}}_i,
      \hat{\boldsymbol{X}}_j\right) \\
      &- \frac{2}{nm} \sum_i^n \sum_j^m \kappa \left(\hat{\boldsymbol{X}}_i,
      \hat{\boldsymbol{Y}}_j\right)
      + \frac{1}{m(m-1)} \sum_{j\neq i} \kappa \left(\hat{\boldsymbol{Y}}_i,
      \hat{\boldsymbol{Y}}_j\right)
  \end{align*}
  where $\kappa$ is some radial basis kernel. Suppose that $m, n \to \infty$ and
  $m/(m+n) \to \rho \in (0, 1)$. Then under the null hypothesis of $F = G \circ
  \boldsymbol{W}$,
  \begin{align*}
    \vert T_{n, m} (\hat{\boldsymbol{X}}, \hat{\boldsymbol{Y}})
    - T_{n, m} (\boldsymbol{X}, \boldsymbol{Y}\boldsymbol{W}) \vert \overset{a.s.}\to 0,
  \end{align*}
  and $\vert T_{n, m} (\boldsymbol{X}, \boldsymbol{Y}\boldsymbol{W}) \vert \to
  0$ as $n, m \to \infty$, where $\boldsymbol{W}$ is any orthogonal matrix such
  that $F = G \circ \boldsymbol{W}$. In addition, under the alternative
  hypothesis $F \neq G \circ \boldsymbol{W}$ for any orthogonal matrix
  $\boldsymbol{W} \in \mathbb{R}^{d \times d}$ that is dependent on $F$ and $G$
  but independent of $m$ and $n$, we have
  \begin{align*}
    \vert T_{n, m} (\hat{\boldsymbol{X}}, \hat{\boldsymbol{Y}})
    - T_{n, m} (\boldsymbol{X}, \boldsymbol{Y}\boldsymbol{W}) \vert \overset{a.s.}\to 0,
  \end{align*}
  and $\vert T_{n, m} (\boldsymbol{X}, \boldsymbol{Y}\boldsymbol{W}) \vert \to c
  > 0$ as $n, m \to \infty$.
\end{theorem}
Simply said, the authors propose using a test statistic that is a kernel-based
function of the latent position estimates obtained from the ASE and show that
it converges to the test statistic obtained using the true but unknown latent
positions under both null and alternative hypotheses.

Together with the work of \cite{gretton-survey} on the use of maximum mean
discrepancy for testing the equivalence of distributions, this result offers an
asymptotically valid and consistent test. Formally, this means that for two
arbitrary but fixed distributions $F$ and $G$, $T_{n,m}(\hat{\boldsymbol{X}},
\hat{\boldsymbol{Y}}) \to 0$ as $n, m \to \infty$ if and only if $F = G$ (up to
$\boldsymbol{W}$). Such a result requires appropriate conditions on the kernel
function $\kappa$ which are satisfied when $\kappa$ is a Gaussian kernel,
$\kappa_g$, defined as
\begin{align*}
  \kappa_{g} \left( \boldsymbol{t} , \boldsymbol{t}' \right)
  & = \exp \left(- \frac{ \left\lVert \boldsymbol{t} - \boldsymbol{t}' \right\rVert_2^2}
    {2 \sigma ^2} \right)
\end{align*}
with any fixed bandwidth $\sigma^2$ \citep{hsbm}.

The intuition behind the maximum mean discrepancy two-sample test is the
following. Under some conditions, the population difference between the average
values of the kernel within and between two distributions is zero if and only
if the two distributions are the same. Hence, using a sample test statistic
that is consistent for the this difference and rejecting for the large values
thereof leads to a consistent test.

No closed form of the finite-sample distribution of this test statistic is
known, for graphs or in the general setting, so it is not immediately clear how
to calculate the critical value given a significance level $\alpha$. The authors
of \cite{nonpar} propose using permutation resampling in order to approximate
the distribution of the test statistic under the null. The permutation version
of the test is computationally expensive, but practically feasible. Alternatives
to the permutation test include using a $\chi^2$ asymptotic approximations
\citep{gretton-survey}.

\section{Correcting the nonvalidity of the test}
\label{sec:fix}
\subsection{Source of the nonvalidity}
The limiting result in the previous section should, however, be taken with
caution for graphs of finite order. Even though the ASE estimates converge to
the true latent positions, and the test statistic using the estimates converges
to the one using the true values, for any finite $n$ and $m$ there is still
variability associated with these estimates as described by Theorem
\ref{rdpg_clt}.

When the graphs are of the same order, the variability introduced by the
estimates instead of the true latent positions is the same for the two graphs.
Hence, the two embeddings have equal distributions under the null hypothesis, up
to orthogonal nonidentifiability. This leads to a valid and consistent test, as
demonstrated experimentally in both \cite{nonpar} and our Section
\ref{sec:experiments}.

However, recall that the approximate finite-sample distribution of the ASEs has
variance that depends on the number of vertices. Suppose that we have a graph of
order $n$, with adjacency matrix $\boldsymbol{A}$ generated according to
$(\boldsymbol{A}, \boldsymbol{X}) \sim RDPG(F, n)$ and a graph of order $m$,
with adjacency matrix $\boldsymbol{B}$ generated according $(\boldsymbol{B},
\boldsymbol{Y} ) \sim RDPG(G, m)$. From the central limit result stated above,
the distributions of the ASEs of the two graphs, conditioned on the true latent
positions, are
\begin{align}
  \hat{\boldsymbol{X}}_i \vert \boldsymbol{X}_i \overset{approx}{\sim} \mathcal{N}
  \left(\boldsymbol{X}_i \boldsymbol{W}_{\boldsymbol{X}},
  \frac{\boldsymbol{\Sigma}(\boldsymbol{X}_i)}{n} \right)
  \quad \mathrm{and} \quad
  \hat{\boldsymbol{Y}}_i \vert \boldsymbol{Y}_i \overset{approx}{\sim} \mathcal{N}
  \left(\boldsymbol{Y}_i \boldsymbol{W}_{\boldsymbol{Y}},
  \frac{\boldsymbol{\Sigma}(\boldsymbol{Y}_i)}{m} \right), \label{distributions-ase}
\end{align}
where $\boldsymbol{W}_{\boldsymbol{X}}$ and $\boldsymbol{W}_{\boldsymbol{Y}}$
are orthogonal matrices present due to the model-based orthogonal
nonidentifiablity. The unconditioned distributions of the ASEs are not equal
whenever $m\neq n$, even if $\boldsymbol{X}_i$ and $\boldsymbol{Y}_i$ have the
same distribution, i.e. even if $F = G$. Thus, as long as the graphs are not of
the exact same order, the collection $\left\{ \hat{\boldsymbol{X}}_1, \hdots,
  \hat{\boldsymbol{X}}_n, \hat{\boldsymbol{Y}}_1, \hdots \hat{\boldsymbol{Y}}_m
\right\}$ is not exchangeable under the null hypothesis, even up to orthogonal
nonidentifiability. This places the distributions of the ASEs of two graphs of
different order in the alternative of the kernel-based test of
\cite{gretton-survey}, despite the fact that the distributions of the true
latent positions would fall under the null. In many cases, the subsequent
kernel-based test is sensitive enough to pick up these differences in
distributions, which makes the size of the test grow as the sample sizes diverge
from each other.

\begin{figure}[t]
    \begin{center}
    \includegraphics[width=0.75\textwidth]{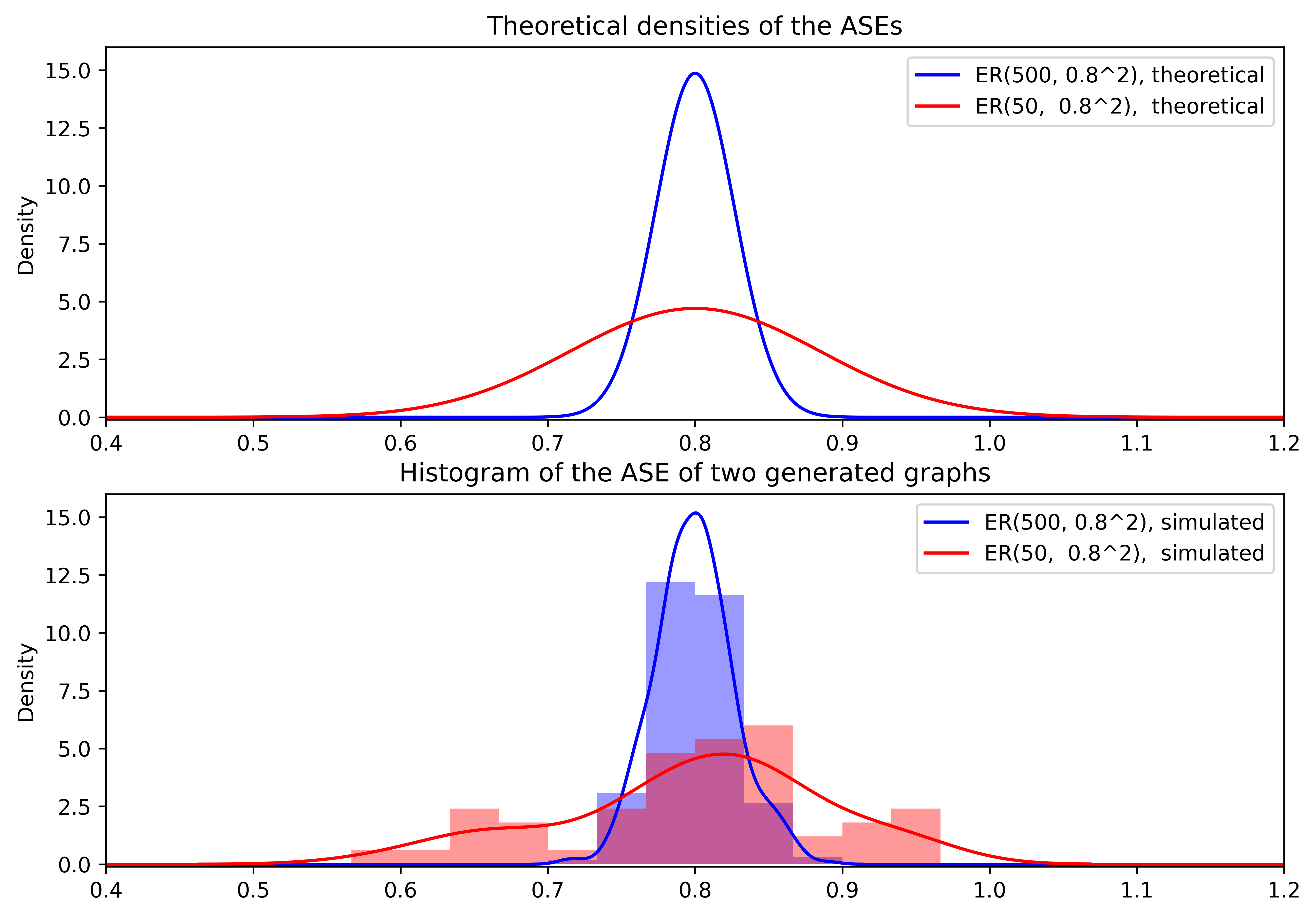}
    \end{center}
    \caption{ A visualization of the ASEs for the Erd\"{o}s-R\'{e}nyi graphs
      with the same edge probability, but vastly different orders. Top:
      theoretical densities of the ASEs; bottom: the histogram of the
      ASEs of two generated graphs, with kernel density estimates. }
    \label{fig:nonpar-nonvalidity}
\end{figure}

Consider the following simple example. Suppose that the graphs have
distributions $\boldsymbol{A} \sim ER(n, p^2)$ and $\boldsymbol{B} \sim ER(m,
p^2)$. Then, the distributions of the ASEs become
\begin{align}
  \hat{\boldsymbol{X}}_i \overset{approx}{\sim} \mathcal{N}
  \left(p, \frac{1 - p^2}{n} \right)
  \quad \mathrm{and} \quad
  \hat{\boldsymbol{Y}}_i \overset{approx}{\sim} \mathcal{N}
  \left(p, \frac{1-p^2}{m} \right).
\end{align}
up to an orthogonal nonidentifiablity, which in a single dimension is just a sign
flip. 

A visualization of this specific case with parameters $n=500, m=50$, and $p=0.8$
is provided in Figure \ref{fig:nonpar-nonvalidity}. The ASEs have substantially
different distributions from each other, despite the identical distributions of
the true latent positions. As will be demonstrated in Section
\ref{sec:experiments}, in this case the nonparametric test developed by
\cite{gretton-survey} and employed by \cite{nonpar} rejects more often than the
significance level $\alpha$, as it should.

Indeed, the test of \cite{gretton-survey} cannot be used directly on the
adjacency spectral embeddings of two graphs of different order to test for the
equivalence of the distributions of the latent positions, as it is not valid.

\subsection{Corrected adjacency spectral embeddings}
\begin{figure}[t]
  \begin{center}
    \includegraphics[width=0.85\textwidth]{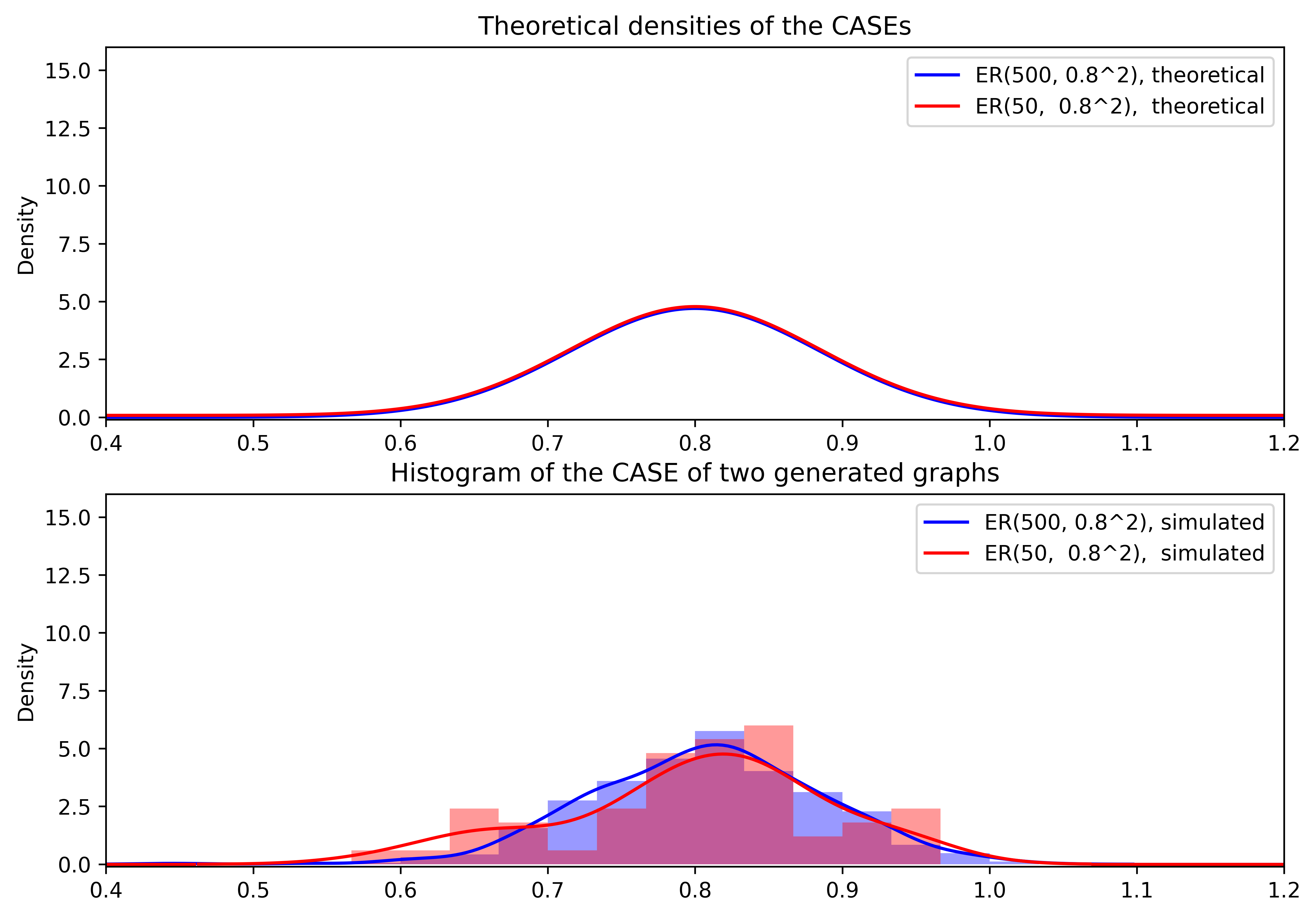}
  \end{center}
  \caption{ A visualization of the CASEs for the Erd\"{o}s-R\'{e}nyi graphs with
    the same edge probability, but vastly different orders. Top: theoretical
    densities of the corrected ASEs; bottom: the histogram of the corrected ASEs
    of two generated graphs, with kernel density estimates. }
  \label{fig:nonpar-validity}
\end{figure}
We propose modifying the adjacency spectral embeddings of one of the graphs by
injecting appropriately scaled Gaussian noise. The noise inflates the variances
of the ASE of the larger graph to approximately the same value as the smaller
graph and makes the latent positions exchangeable under the null hypothesis.

\begin{definition}[Corrected Adjacency Spectral Embedding]
  Consider two $d$-dimensional random dot product graphs $(\boldsymbol{A},
  \boldsymbol{X}) \sim RDPG(F, n)$ and $(\boldsymbol{B}, \boldsymbol{Y} ) \sim
  RDPG(G, m)$. Without loss of generality, assume that $n > m$. For every row in
  the adjacency spectral embedding of the larger graph,
  $\hat{\boldsymbol{X}}_i$, consider estimating its variance using the plug-in
  estimator from Equation \ref{plug-in-estimator}, and then sampling a point
  $\boldsymbol{\epsilon}_{\hat{\boldsymbol{X}}_i} \sim \mathcal{N}(\mathbf{0},
  \left(\frac{1}{m} - \frac{1}{n} \right)
  \hat{\boldsymbol{\Sigma}}(\hat{\boldsymbol{X}}_i))$. For every row in the
  adjacency spectral embedding of the smaller graph, $\hat{\boldsymbol{Y}}_j$,
  define $\boldsymbol{\epsilon}_{\hat{\boldsymbol{Y}}_j} := \mathbf{0}$. Let
  $\widetilde{\boldsymbol{X}}_i = \hat{\boldsymbol{X}}_i +
  \boldsymbol{\epsilon}_{\hat{\boldsymbol{X}}_i}$ for all $i$ and
  $\widetilde{\boldsymbol{Y}}_j = \hat{\boldsymbol{Y}}_j +
  \boldsymbol{\epsilon}_{\hat{\boldsymbol{Y}}_j}$ for all $j$. We denote the
  matries whose rows consist of these new vectors $\widetilde{\boldsymbol{X}}$
  and $\widetilde{\boldsymbol{Y}}$, respectively, and we call them the {\it
    corrected adjacency spectral embeddings (CASE)}. The corrected adjacency
  spectral embeddings of two graphs of the same order are equal to the standard
  adjacency spectral embeddings.
\end{definition}

The motivation for the preceding definition is as follows. Recall that we have
assumed without the loss of generality that $n > m$. Conditioned on the true
latent positions, the rows of the corrected adjeacency spectral embeddings have
distributions that are given by
\begin{equation}
\begin{aligned}
  \widetilde{\boldsymbol{X}}_i \vert \boldsymbol{X}_i
  &\overset{approx}{\sim} \mathcal{N}
  \left(\boldsymbol{X}_i \boldsymbol{W}_{\boldsymbol{X}},
  \frac{\boldsymbol{\Sigma}(\boldsymbol{X}_i)}{n}
  + \left(\frac{1}{m} - \frac{1}{n} \right)
  \hat{\boldsymbol{\Sigma}}(\hat{\boldsymbol{X}}_i) \right) \\
  \widetilde{\boldsymbol{Y}_i}\vert \boldsymbol{Y}_i
  &\overset{approx}{\sim}  \mathcal{N}
  \left(\boldsymbol{Y}_i \boldsymbol{W}_{\boldsymbol{Y}}, \frac{\boldsymbol{\Sigma}(\boldsymbol{Y}_i)}{m} \right).
\end{aligned}
\end{equation}

Unlike Equation \ref{distributions-ase}, these distributions are approximately
the same, up to orthogonal transformations $\boldsymbol{W}_{\boldsymbol{X}}$ and
$\boldsymbol{W}_{\boldsymbol{Y}}$. This is true regardless of the ratio of graph
orders, as long the true latent positions $\boldsymbol{X}_i, \boldsymbol{Y}_i$
have the same distribution and $\hat{\boldsymbol{\Sigma}}$ is a good estimator
of $\boldsymbol{\Sigma}$. As an illustrative example, we revisit the ER
ilustration from the previous section. A visualization of the theoretical and
simulated CASEs of two ER graphs with vastly different orders is presented in
Figure \ref{fig:nonpar-validity}. Both the theoretical and the simulated
corrected embeddings have the same distribution. Hence, the corrected adjacency
spectral embeddings can be used as inputs to the latent distribution test of
\cite{nonpar}.

We note that due to the exact equivalence of the maximum mean discrepancy test
of \cite{gretton-survey}, the Energy distance two-sample test \cite{energy}, the
Hilbert-Schmidt independence criterion \citep{hsic}, and distance correlation
\citep{dcorr, dcorr-unbiased} test for independence, any of these four can be
used as a subsequent test interchangeably \citep{exact-equivalence-1,
  exact-equivalence-2}. In the case of the latter two of the four, one first has
to concatenate the two embeddings, define an auxiliary label vector, and then
perform the independence test. For more on this procedure, sometimes called
$k$-sample transform, see \cite{exact-equivalence-1}.

It may also be possible to use other independence tests framed as two-sample
tests to test for the equivalence of the latent distributions after the
embeddings have been obtained and corrected. Such tests include, but are not
limited to RV \citep{rv-1, rv-2} which is the multivariate generalization of the
Pearson correlation \citep{pearson}, canonical component analysis \citep{cca},
and multiscale graph correlation \citep{mgc-1, mgc-2}. The power of the
multiscale graph correlation against some alternatives has been studied in the
graph setting in \cite{varjavand}. However, no theoretical guarantees, at least
known to us, have been established in the graph setting for any of these tests.

\section{Simulation study}
\label{sec:experiments}
We conduct a simulation study comparing the latent distribution tests that use
regular and corrected ASEs. We use graphs generated from the ER, SBM and RDPG
models in our experiments. However, we always estimate the variances of the ASE
using the generic plug-in estimator for the RDPG model, provided in Equation
\ref{plug-in-estimator}. That is, we do not use the knowledge that the latent
distribution is truly a point-mass, or a mixture thereof, anywhere in our
experiments.

The implementation of the latent distribution test used in this simulation study
is incorporated into {\bf graspologic} \citep{graspy} Python package, both for
ASE and CASE. This implementation exploits the exact equivalence with
independence tests described above. Code that is compatible with the latest
version of {\bf graspologic} and can be used to reproduce all of the simulations is
available at
\href{https://github.com/alyakin314/correcting-nonpar}{https://github.com/alyakin314/correcting-nonpar}.

We set the number of permutations used to generate the null distribution
to 200.
For a task like this, it is quite common to use a gaussian kernel with a
bandwith selected using a median heuristic \citep{median-heuristic}, which in
practice might be more sensetive than most arbitrarily chosen constant
bandwidths. However since the theoretical result holds only a fixed kernel, we
chose to use a Gaussian kernel with a fixed bandwidth $\sigma = 0.5$ throughout
our experiments.

\subsection{Erd\"{o}s-R\'{e}nyi Graphs: Validity and Consistency}
\begin{figure}[t]
  \begin{center}
    \includegraphics[width=0.49\textwidth]{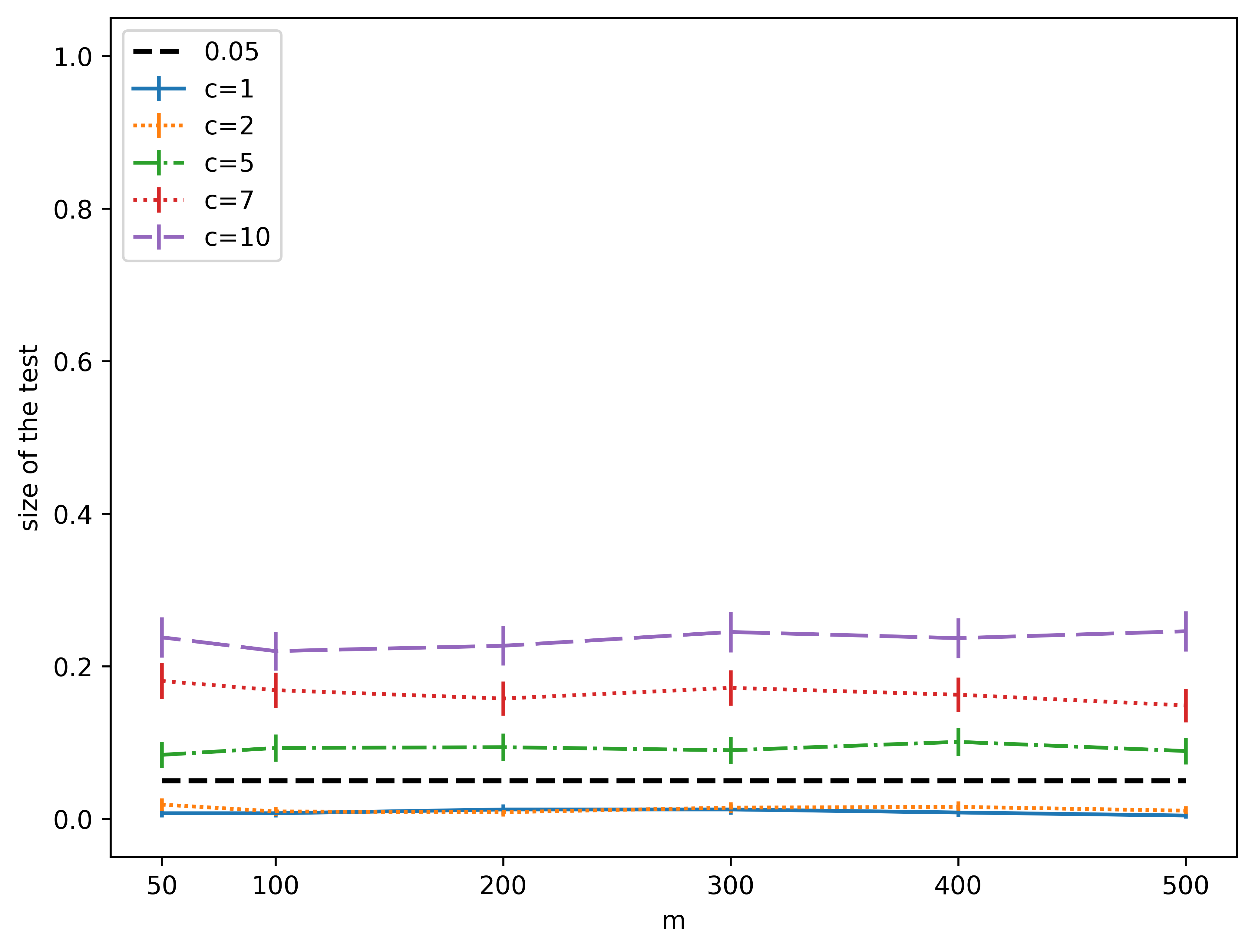}
    \includegraphics[width=0.49\textwidth]{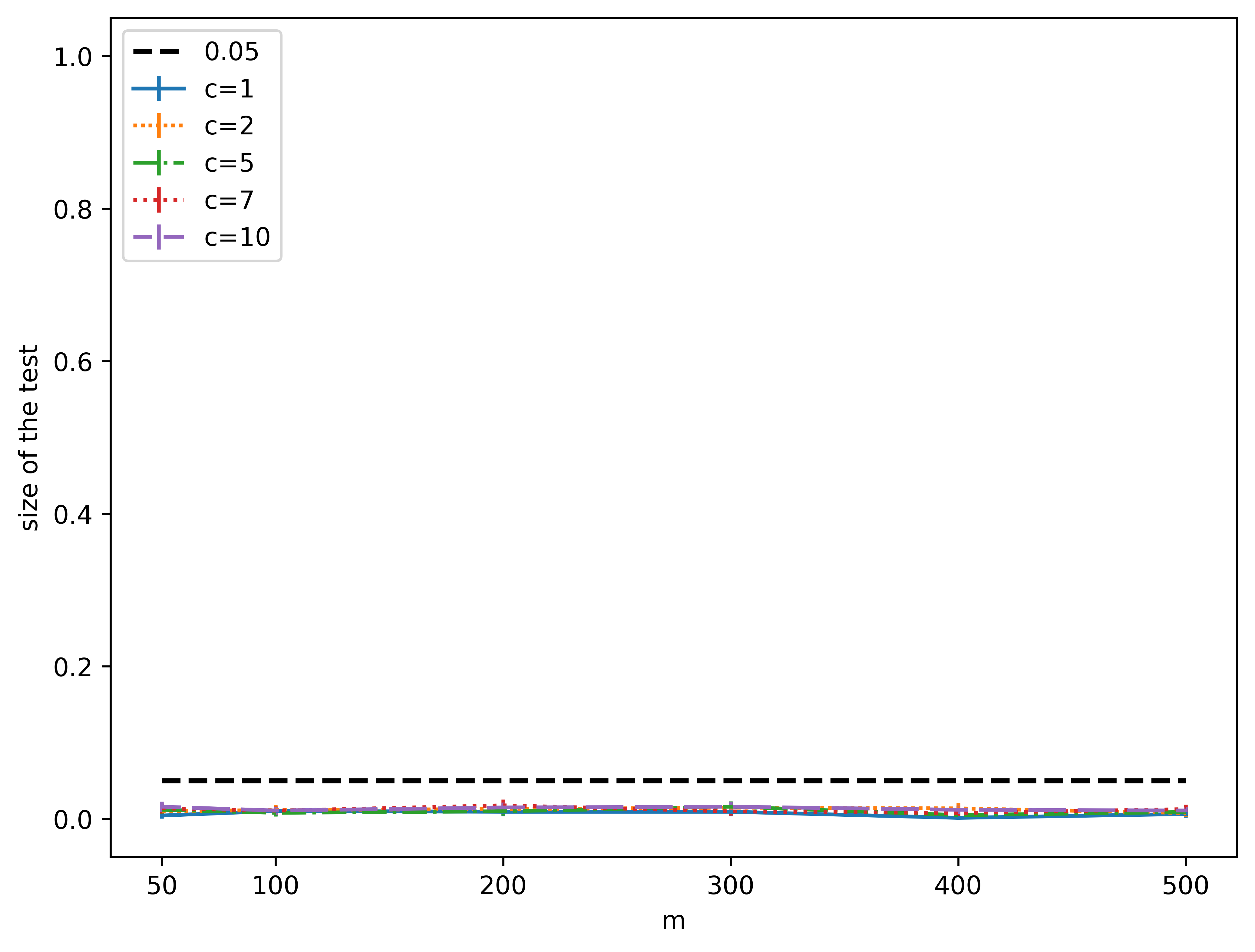}
  \end{center}
  \caption{Size of the nonparametric latent distribution permutation tests that
    use the standard ASE (left) and the CASE (right). Graphs are $\boldsymbol{A}
    \sim ER(n, 0.8^2)$ and $\boldsymbol{B} \sim ER(cn, 0.8^2)$. Error bars
    represent 95\% confidence interval.}
  \label{fig:nonpar-size}
\end{figure}
We generate pairs of graphs from the null hypothesis of the test:
$\boldsymbol{A} \sim ER(n, p^2)$ and $\boldsymbol{B} \sim ER(m, p^2)$ with $m =
cn$. We consider different ratios of the graphs orders $c \in \{1, 2, 5, 7, 10,\}$,
and different smaller graph orders $n \in \{50, 100, 200, 300, 400, 500
\}$. We use the latent position $p = 0.8$, which corresponds to the
Erd\"{o}s-R\'{e}nyi graphs with the edge probability of $0.64$. We always embed
the graphs into one dimension and we overcome orthogonal nonidentifiability by
flipping the signs of the ASE of a graph if their median is negative. $1000$
Monte-carlo replications are used for each of combination of $c$ and $n$ tested.

We set $\alpha $ to $0.05$ and report the sizes of the test in
Figure \ref{fig:nonpar-size}. The size of the test that use the standard ASE
grows as a function of $c$ rendering it invalid for graphs of different
sizes. The size of the test that uses the CASE remains below $0.05$ across all
choices of $c$ and $n$ considered.

In general, the size of the permutation tests should be exactly $\alpha$.
However, due to the intricate dependence behavior of the graph spectral
embeddings \citep{avanti-clt, minh-spectral}, the tests ends up being
conservative. The extent to which the test is conservative is dependent on the
model from which the graphs were generated, and thus cannot be easily corrected.
The scope of this work is limited to correcting the invalidity phenomenon and
not the conservatism of this test.

\begin{figure}[t]
  \begin{center}
    \includegraphics[width=0.75\textwidth]{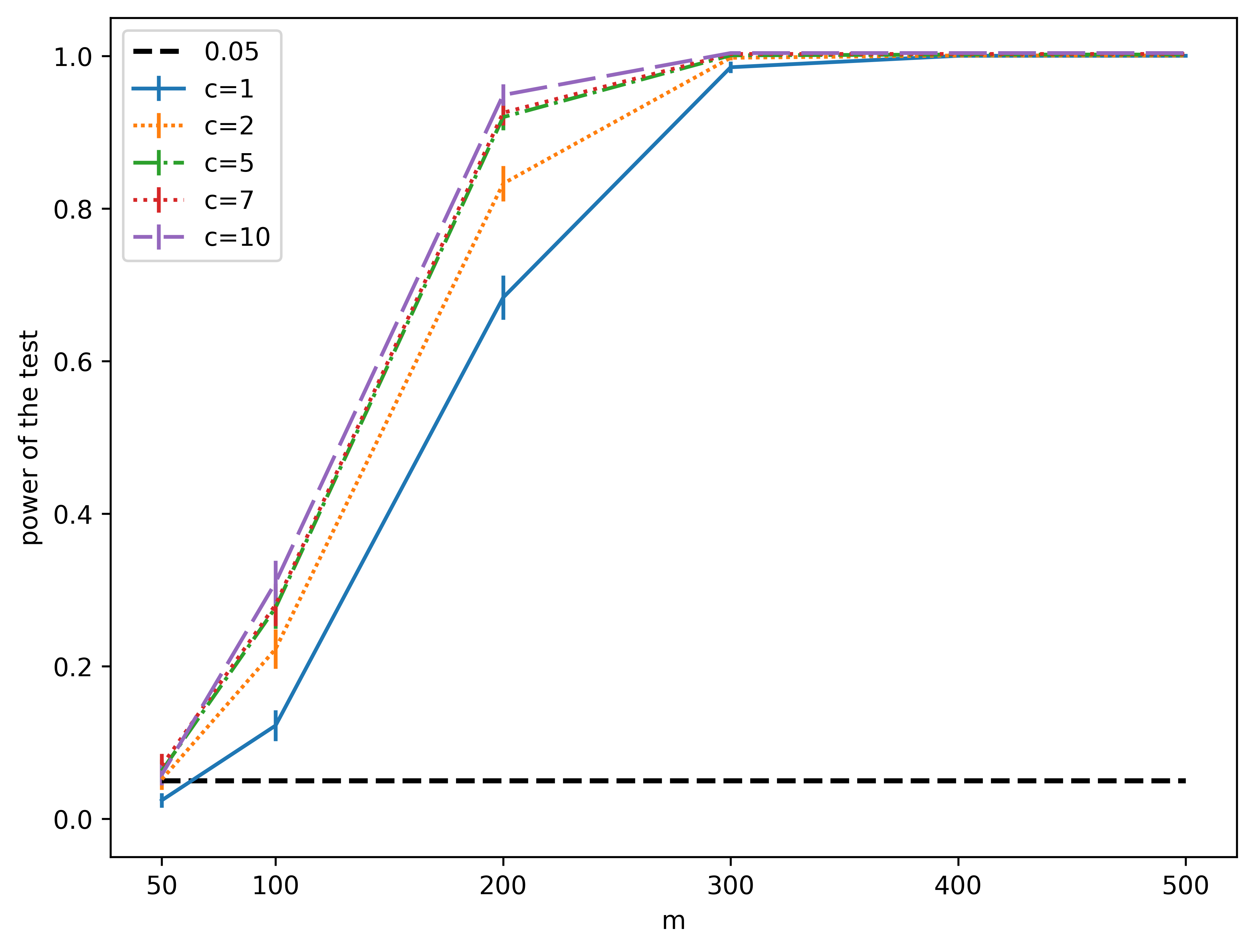}
  \end{center}
  \caption{Power of the nonparametric latent distribution permutation test that
    uses CASE against the alternative with graphs generated from $\boldsymbol{A}
    \sim ER(n, 0.8^2)$ and $\boldsymbol{B} \sim ER(cn, 0.79^2)$. Error bars
    represent 95\% confidence interval.}
  \label{fig:nonpar-power}
\end{figure}
We also study the behavior of the test under the alternative hypothesis in order
to assess its power. We use the alternative hypothesis $\boldsymbol{A} \sim
ER(n, p^2)$ and $\boldsymbol{B} \sim ER(m, q^2)$, with $p=0.8$ and $q=0.79$ and
$m = cn$ for various ratios $c$. We again consider the graph order ratios $c \in
\{1, 2, 5, 7, 10\}$, and smaller graph orders $n \in \{50, 100, 200, 300,
400, 500\}$. For $c = 1$, CASE overlaps exactly with the standard ASE, so the
testing procedure is the same as the original test of \cite{nonpar}. For all
other choices of $c$, the  original test is not valid, and is thus omitted
from study.

The results of this study are presented in Figure \ref{fig:nonpar-power}. The
power of the test goes to one as the sample size increases for all choices of
$c$ used, which suggests that the test that uses CASE is still consistent. We
note that for any given $n$, the power of the test grows as $c$ grows; this
behavior is expected, since the number of vertices in one graph is held constant
and the number of vertices in the other increases, so the total number of
observations grows.

\subsection{Stochastic Block Model Graphs: Higher Dimensions}
\begin{figure}[t]
  \begin{center}
    \includegraphics[width=0.49\textwidth]{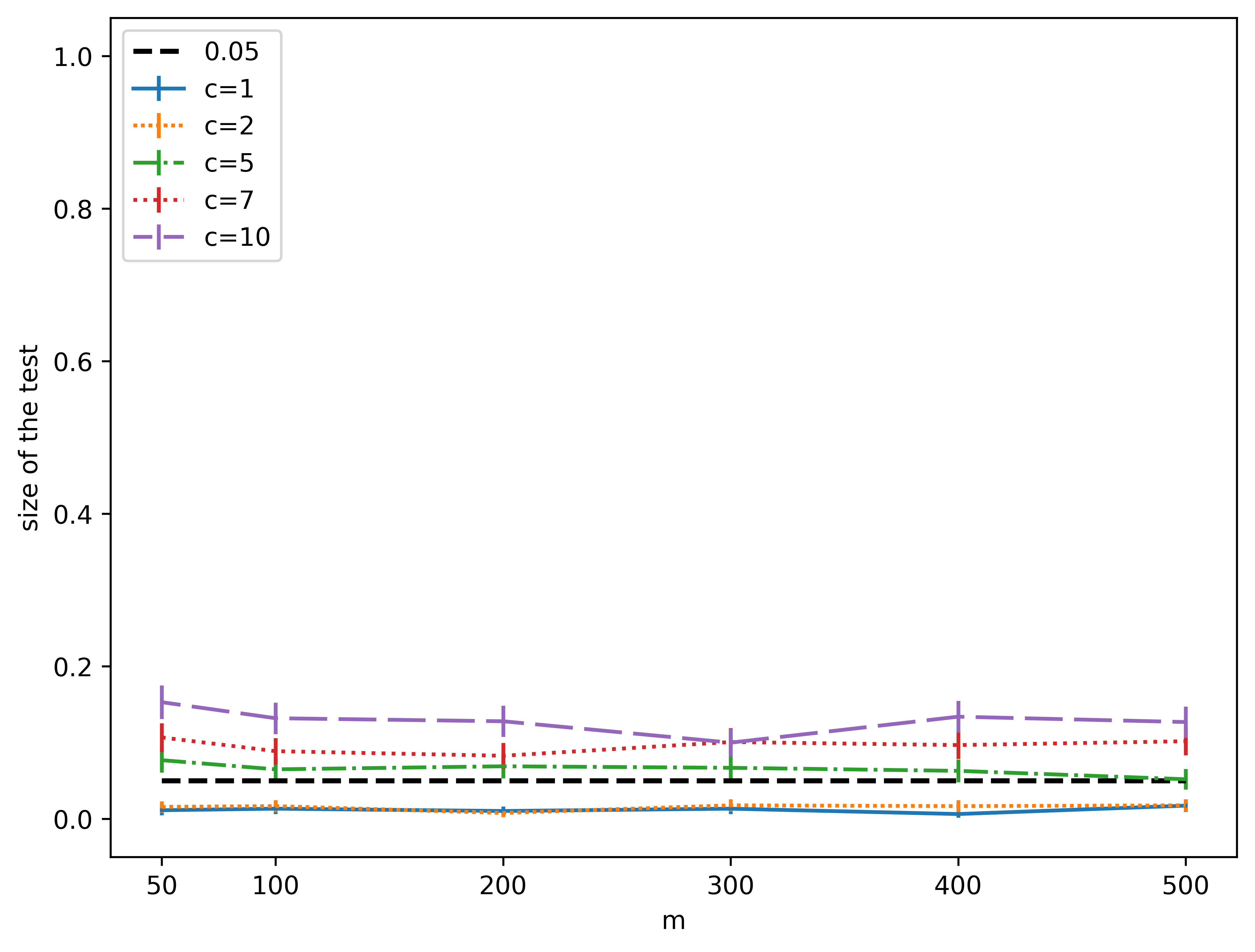}
    \includegraphics[width=0.49\textwidth]{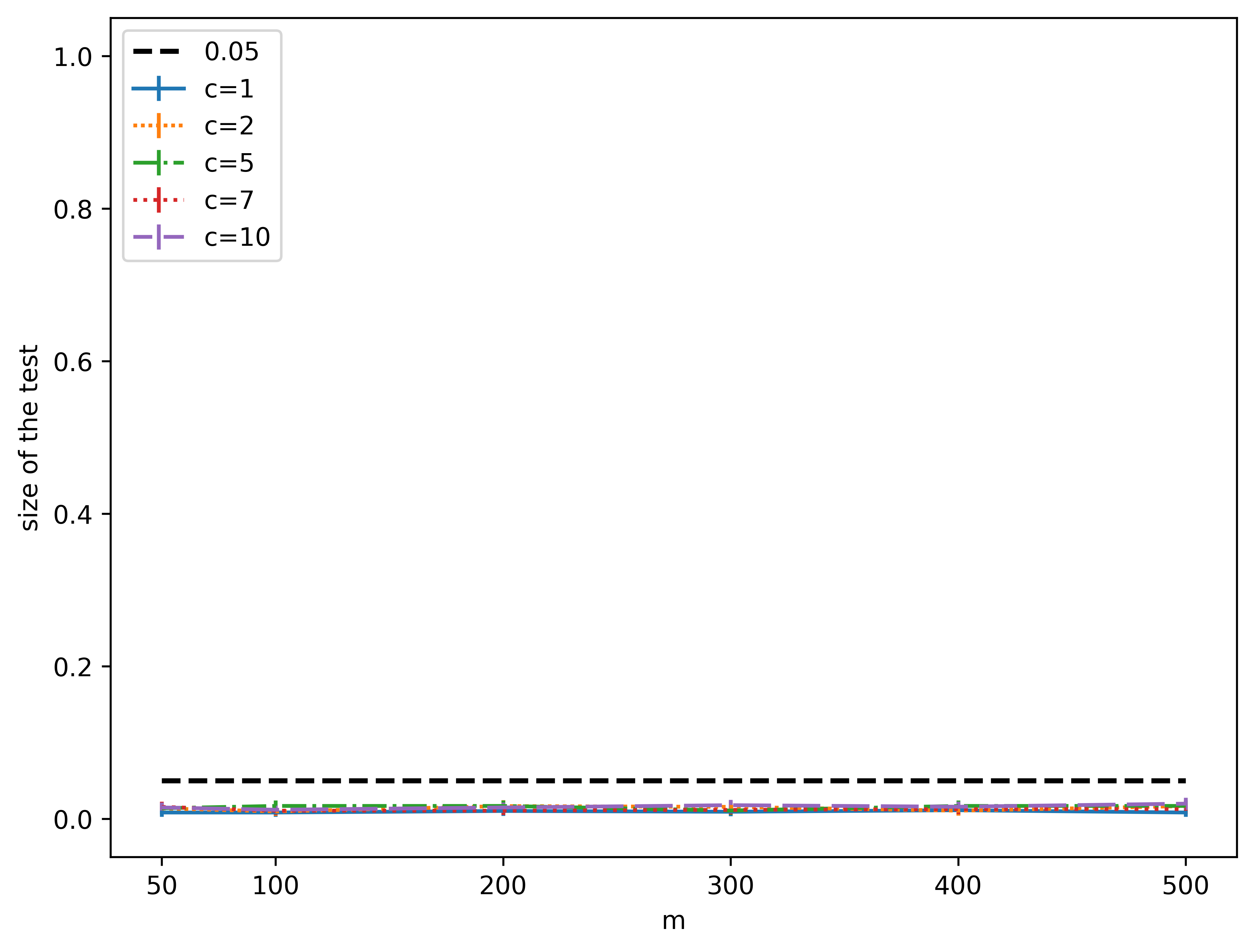}
  \end{center}
  \caption{Size of the nonparametric latent distribution permutation tests that
    use the regular ASE (left) and the CASE (right). Graphs are $\boldsymbol{A}
    \sim SBM (n, \boldsymbol{\pi}, \boldsymbol{P})$ and $\boldsymbol{B} \sim SBM
    (cn, \boldsymbol{\pi}, \boldsymbol{P})$. Error bars represent 95\% confidence
    interval.}
  \label{fig:nonpar-size-sbm}
\end{figure}
We repeat the validity and consistency experiments, but use $3$-block SBMs,
instead of ER graphs. In all simulations we use the vector of prior
probabilities $\boldsymbol{\pi} = [0.4, 0.3, 0.3]^T$. To estimate size, we use
graphs $\boldsymbol{A} \sim SBM (n, \boldsymbol{\pi}, \boldsymbol{P})$ and
$\boldsymbol{B} \sim SBM (m, \boldsymbol{\pi}, \boldsymbol{P})$, where the
block-probability matrix $\boldsymbol{P} = \boldsymbol{Z} \boldsymbol{Z}^T$
is obtained using the matrix of latent positions $\boldsymbol{Z}$ is being
parametrized by spherical coordinates
\begin{align*}
  \boldsymbol{Z}
  &= \begin{bmatrix}
       \begin{bmatrix}
         r \sin(\boldsymbol{\theta}_1) \sin(\boldsymbol{\omega}_1) \\
         r \sin(\boldsymbol{\theta}_1) \cos(\boldsymbol{\omega}_1) \\
         r \cos(\boldsymbol{\theta}_1)
       \end{bmatrix},
       \begin{bmatrix}
         r \sin(\boldsymbol{\theta}_2) \sin(\boldsymbol{\omega}_2) \\
         r \sin(\boldsymbol{\theta}_2) \cos(\boldsymbol{\omega}_2) \\
         r \cos(\boldsymbol{\theta}_2)
       \end{bmatrix},
       \begin{bmatrix}
         r \sin(\boldsymbol{\theta}_3) \sin(\boldsymbol{\omega}_3) \\
         r \sin(\boldsymbol{\theta}_3) \cos(\boldsymbol{\omega}_3) \\
         r \cos(\boldsymbol{\theta}_3)
       \end{bmatrix}
    \end{bmatrix}^T,
  \end{align*}
  where $r = 0.9$, $\boldsymbol{\theta} = [0, 0.2, 0.4, 0.5]^T$, and
  $\boldsymbol{\omega} = [0.00, 0.10, 0.05, 0.05]^T$ (the fourth coordinate will
  become relevant for the evaluation of power). Numerically,
  \begin{align*}
    \boldsymbol{Z}
    &\approx \begin{bmatrix}
               \begin{bmatrix}
                 0.000 \\ 0.000 \\ 0.900
               \end{bmatrix},
               \begin{bmatrix}
                 0.018 \\ 0.178 \\ 0.882
               \end{bmatrix},
               \begin{bmatrix}
                 0.018 \\ 0.350 \\ 0.829
               \end{bmatrix}
             \end{bmatrix}^T
  \end{align*}
  and 
\begin{align*}
  \boldsymbol{P}
  &\approx \begin{bmatrix}
    0.810 & 0.794 & 0.746 \\
    0.794 & 0.810 & 0.794 \\
    0.746 & 0.794 & 0.810
  \end{bmatrix}.
  \end{align*}

  Exactly as the one-dimensional case, we constrain $m = cn$ and consider the
  graph order ratios $c \in \{1, 2, 5, 7, 10\}$, and smaller graph orders $n \in
  \{50, 100, 200, 300, 400, 500\}$. We always embed into the true dimension
  $d=3$. We overcome orthogonal nonidentifiability by aligning the medians of
  the embeddings to be in the same quadrant by flipping all of the signs on one
  of them if they do not match. The size of the tests at $\alpha = 0.05$ is
  presented in Figure \ref{fig:nonpar-size-sbm}. Similarly to the
  one-dimensional setting, the size of the test that uses standard ASE grows as
  a function of $c$, but is unaffected for the test that uses CASE.
\begin{figure}[t]
  \begin{center}
    \includegraphics[width=0.75\textwidth]{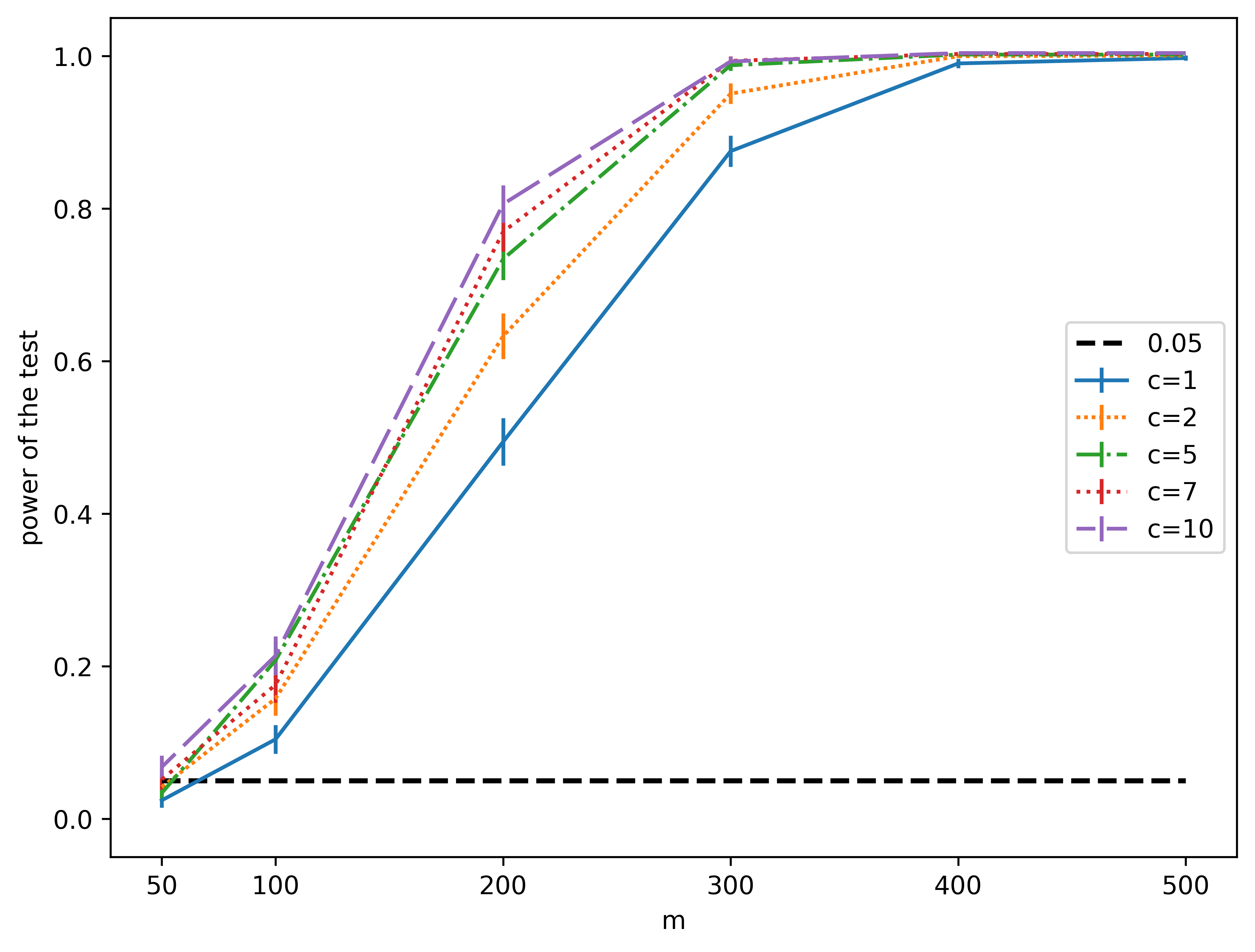}
  \end{center}
  \caption{Power of the nonparametric latent distribution permutation test that
    uses CASE against the alternative with graphs generated from $\boldsymbol{A}
    \sim SBM (n, \boldsymbol{\pi}, \boldsymbol{P})$ and $\boldsymbol{B} \sim SBM
    (m, \boldsymbol{\pi}, \boldsymbol{P}')$,. Error bars represent 95\%
    confidence interval.}
  \label{fig:nonpar-power-sbm}
\end{figure}

To estimate power and demonstrate consistency in higher dimensions we use a pair
of graphs $A$ and $B$, generated from $SBM (n, \boldsymbol{\pi},
\boldsymbol{P})$ and $SBM (m, \boldsymbol{\pi}, \boldsymbol{P}')$, respectively,
where $\boldsymbol{P}$ is as defined above, and $\boldsymbol{P'} =
\boldsymbol{Z'} \boldsymbol{Z'}^T$ with
\begin{align*}
  \boldsymbol{Z'}
  &= \begin{bmatrix}
       \begin{bmatrix}
         r \sin(\boldsymbol{\theta}_1) \sin(\boldsymbol{\omega}_1) \\
         r \sin(\boldsymbol{\theta}_1) \cos(\boldsymbol{\omega}_1) \\
         r \cos(\boldsymbol{\theta}_1)
       \end{bmatrix},
       \begin{bmatrix}
         r \sin(\boldsymbol{\theta}_2) \sin(\boldsymbol{\omega}_2) \\
         r \sin(\boldsymbol{\theta}_2) \cos(\boldsymbol{\omega}_2) \\
         r \cos(\boldsymbol{\theta}_2)
       \end{bmatrix},
       \begin{bmatrix}
         r \sin(\boldsymbol{\theta}_4) \sin(\boldsymbol{\omega}_4) \\
         r \sin(\boldsymbol{\theta}_4) \cos(\boldsymbol{\omega}_4) \\
         r \cos(\boldsymbol{\theta}_4)
       \end{bmatrix}
    \end{bmatrix}^T.
  \end{align*}
  Numerically,
  \begin{align*}
    \boldsymbol{Z'}
    &\approx \begin{bmatrix}
               \begin{bmatrix}
                 0.000 \\ 0.000 \\ 0.900
               \end{bmatrix},
               \begin{bmatrix}
                 0.018 \\ 0.178 \\ 0.882
               \end{bmatrix},
               \begin{bmatrix}
                 0.022 \\ 0.431 \\ 0.790
               \end{bmatrix}
             \end{bmatrix}^T
  \end{align*}
  and 
\begin{align*}
  \boldsymbol{P'}
  &\approx \begin{bmatrix}
    0.810 & 0.794 & 0.711 \\
    0.794 & 0.810 & 0.774 \\
    0.711 & 0.774 & 0.810
  \end{bmatrix}.
  \end{align*}
Note that the only differing feature of the second graph is the latent position
of the vertices in the third block, and the difference is entirely explained by
the position being slightly further away in the $\theta$ coordiante. The graph
orders and ratios of thereof are identical to the ones used in the validity
simulation. The results are presented in the Figure \ref{fig:nonpar-power-sbm}.
The test that uses the CASE remains consistent even in $d=3$.

\subsection{Random Dot Product Graphs: General Setting}
\begin{figure}[t]
  \begin{center}
    \includegraphics[width=\textwidth]{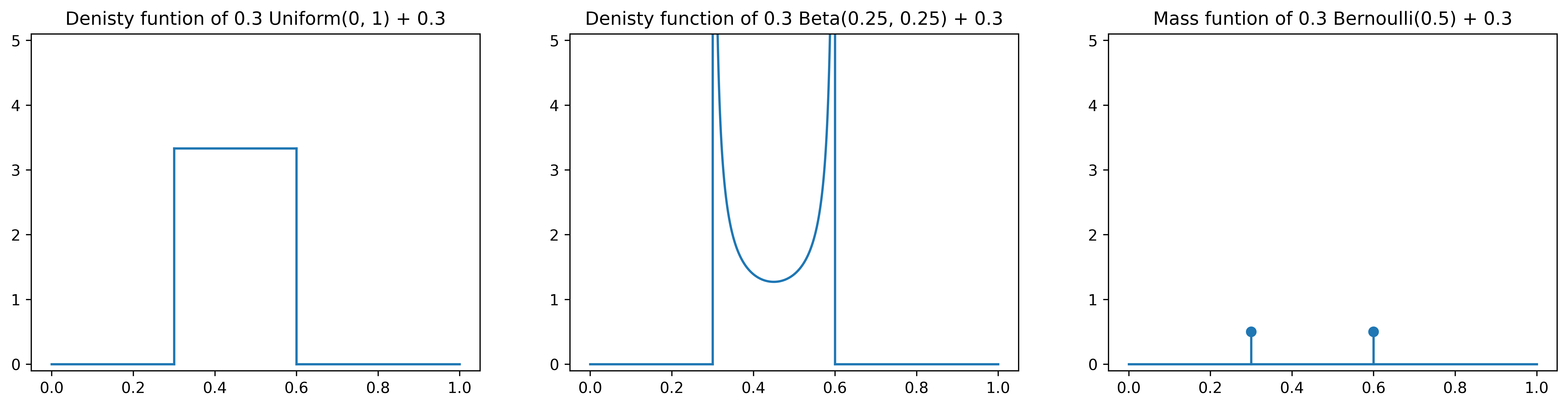}
  \end{center}
  \caption{Density / Mass functions visualization for $F_x= 0.3 \ Uniform(0, 1)
    + 0.3, F_y= 0.3 \ Beta(0.25, 0.25) + 0.3$, and $F_z= 0.3 \ Bernoulli(0.5) +
    0.3$. All three are used as latent position distributions in our
    experiments.}
  \label{fig:rdpg_distributions_visualization}
\end{figure}

Lastly, we present a simulation with continious latent distributions.
Specifically, consider three different distributions:
\begin{align*}
    F_x &= 0.3 \ Uniform(0, 1) + 0.3 \\
    F_y &= 0.3 \ Beta(0.25, 0.25) + 0.3 \\
    F_z &= 0.3 \ Bernoulli(0.5) + 0.3
\end{align*}
Note that all three distributions can be formulated in the context of the $0.3
\ Beta(a, a) + 0.3$ model. Namely, $F_x$ is equivalent to $0.3 \ Beta(1, 1) +
0.3$, $F_y$ is in such a form already, and $0.3 \ Beta(a, a) + 0.3 \to F_z$ as
$a \to 0$. $F_y$ can be thought of as an intermediate step between the $F_x$ and
$F_z$. The visualizations of the density or mass functions of these
distributions are provided in Figure \ref{fig:rdpg_distributions_visualization}.

Also, observe that $F_z$ is nothing more than the latent distribution of a
two-block SBM in a single dimension with a block-probability matrix
\begin{align*}
  \boldsymbol{P}
  &= \begin{bmatrix}
  0.6^2 & (0.6)(0.3) \\
  (0.3)(0.6) & 0.3^2
  \end{bmatrix},
\end{align*}
whereas $F_x$ and $F_y$ can be thought as latent distributions of either DCSBMs
or MMSBMs, as per Remark \ref{rpdg_sbm_equivalence}. Thinking of them as MMSBMs
with $\boldsymbol{Z} = [0.6, 0.3]^T$, the parameter $a$ can be viewed as a
mixing coefficient: $F_x$ has a lot of mixing, $F_y$ has some mixing, and
$F_z$ has two components completely separated.
\begin{figure}[t]
  \begin{center}
    \includegraphics[width=0.49\textwidth]{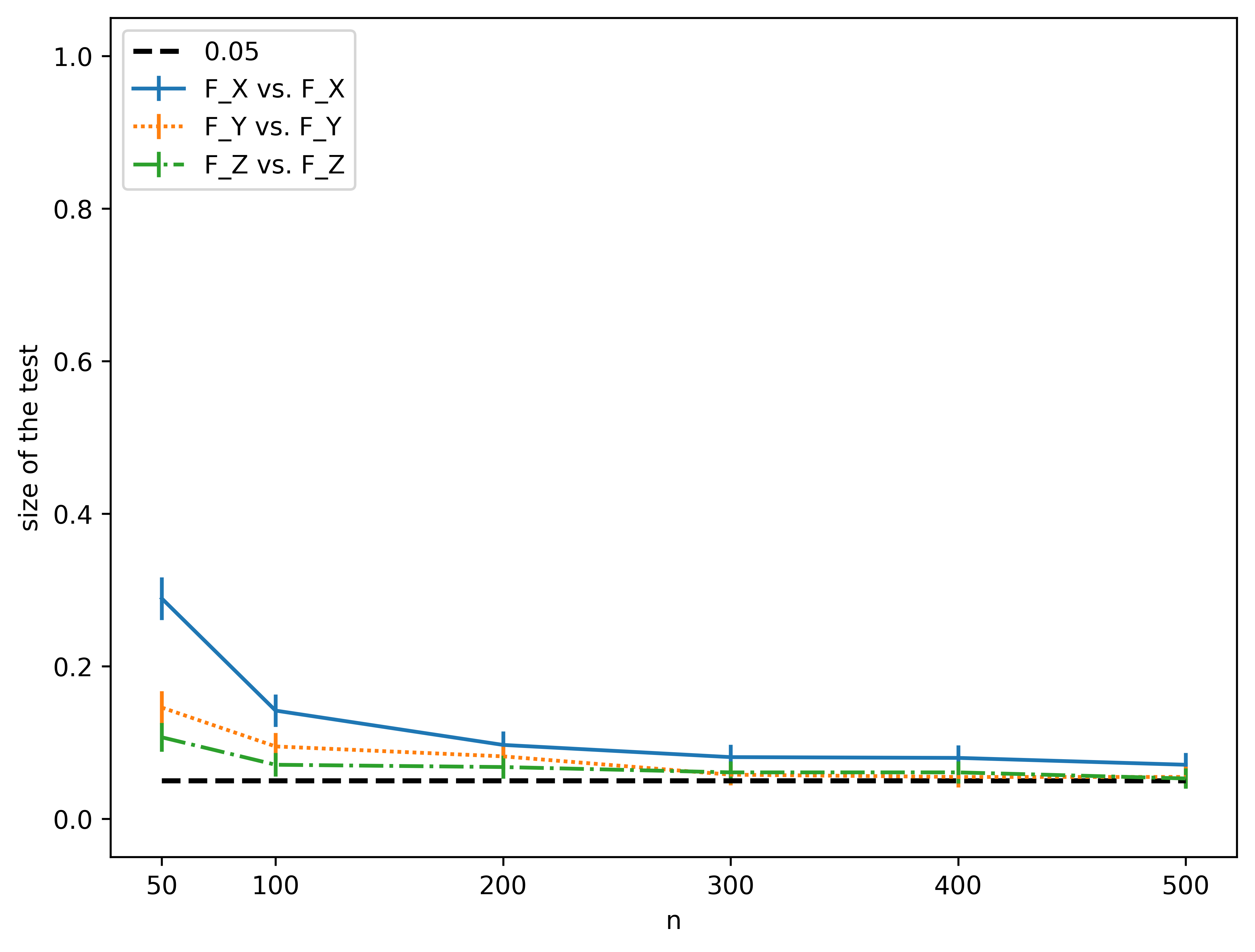}
    \includegraphics[width=0.49\textwidth]{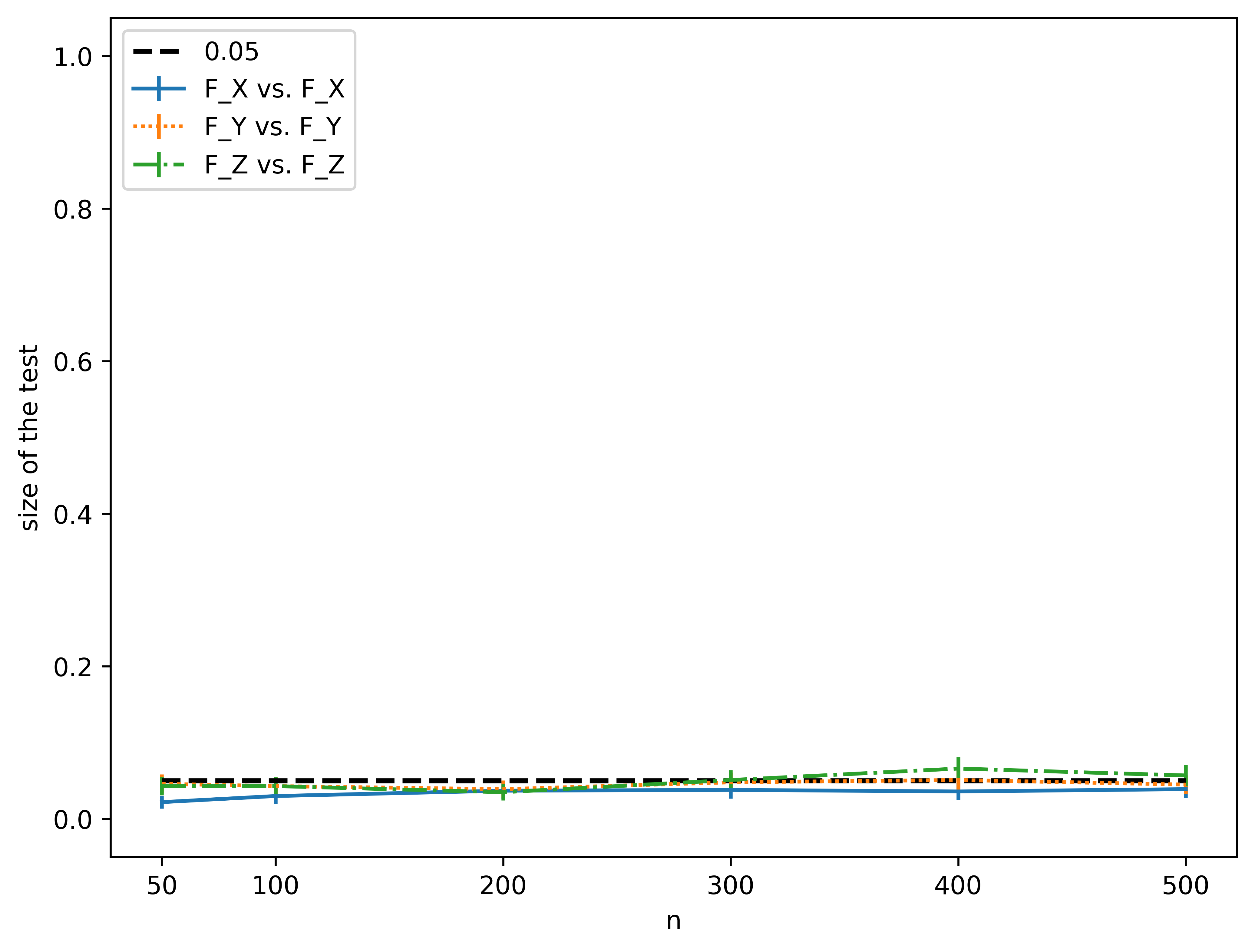}
  \end{center}
  \caption{Size of the nonparametric latent distribution permutation tests that
    use the regular ASE (left) and the CASE (right). Graphs are $\boldsymbol{A}
    \sim RDPG (F, n)$ and $\boldsymbol{B} \sim RDPG (F, m)$, where $F \in \{F_x,
    F_y, F_z\}$. Error bars represent 95\% confidence interval.}
  \label{fig:nonpar-size-rdpg}
\end{figure}

First, we consider graphs $\boldsymbol{A}$ and $\boldsymbol{B}$ generated from
$(\boldsymbol{X}, \boldsymbol{A}) \sim RDPG(F, n)$, and $(\boldsymbol{Y},
\boldsymbol{B}) \sim RDPG(F, m)$, with $m = cn$ This setting is in the null
hypothesis of the latent distribution test. Unlike the previous experiment
settings, we set $c$ to a single value of $10$, and instead vary the
distributions of the latent positions. We consider $F \in \{F_x, F_y, F_z\}$.
The number of vertices of the smaller graph, $n$, is once again varied to be
$\{50, 100, 200, 300, 400, 500\}$. We generate 1000 pairs of graphs for each of
the possible settings, and use both a test that uses ASE and a test that uses
CASE. Like before, we overcome orthogonal nonidentifiability by aligning the
medians of the embeddings via flippting signs.

\begin{figure}[t]
  \begin{center}
    \includegraphics[width=0.49\textwidth]{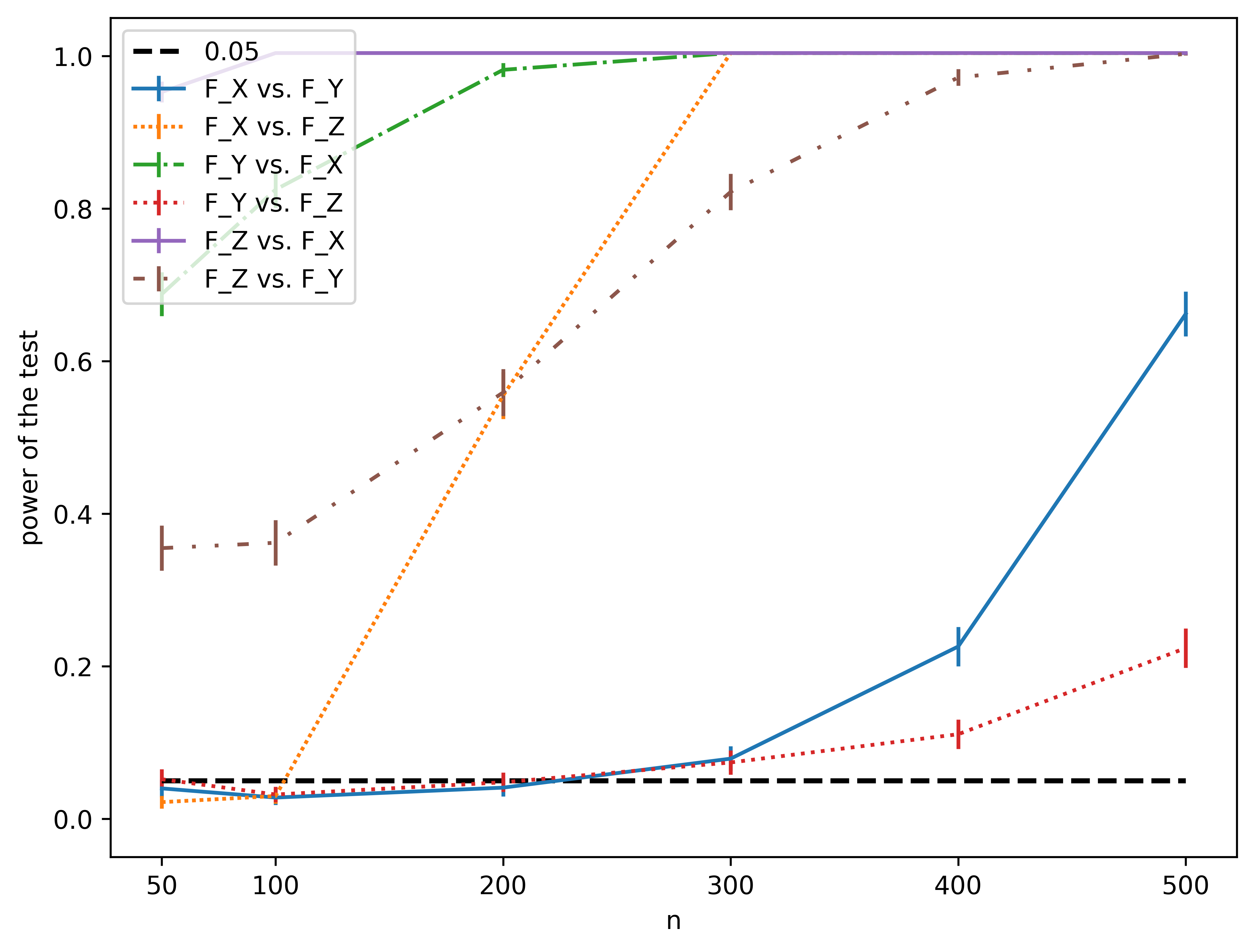}
    \includegraphics[width=0.49\textwidth]{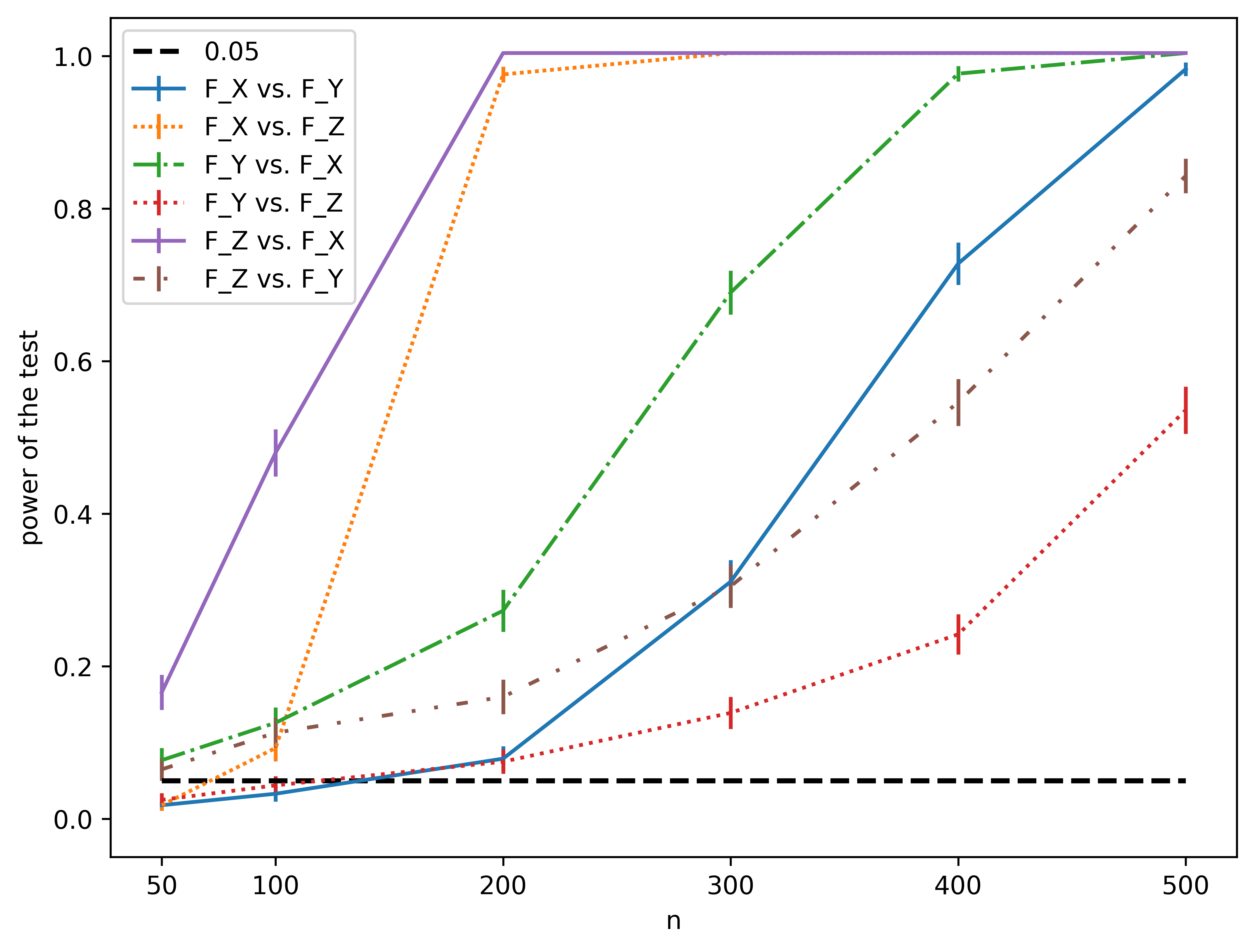}
  \end{center}
  \caption{Power of the nonparametric latent distribution permutation tests that
    use the regular ASE (left) and the CASE (right). Graphs are $\boldsymbol{A}
    \sim RDPG (F_1, n)$ and $\boldsymbol{B} \sim RDPG (F_1, m)$, where $(F_1,
    F_2) \in \{F_x, F_y, F_z\} \times \{F_x, F_y, F_z\}$, except for the cases
    when $F_1 = F_2$, as those are presented in Firgure \ref{fig:nonpar-size-rdpg}.
    Error bars represent 95\% confidence interval.}
  \label{fig:nonpar-power-rdpg}
\end{figure}

Results of this simulation are presented in the Figure
\ref{fig:nonpar-power-rdpg}. Observe that the test that uses ASE is not valid
for all three distributions of the latent positions, which is especially
clear at the small $n$. On the other hand - test that uses CASE remains
experimentally valid for all three distributions: with no mixing, little mixing,
and a lot of mixing.

Next, we simulate the power of the test under different alternatives.
Specifically, we generate pairs of graphs $\boldsymbol{A}$ and $\boldsymbol{B}$,
where $(\boldsymbol{X}, \boldsymbol{A}) \sim RDPG(F, n)$, and
$(\boldsymbol{Y}, \boldsymbol{B}) \sim RDPG(F', m)$, where $m = cn$. We keep the
same settings of $n$ and $c$ as for the study of the test size, and pick $(F,
F')$ from the collection $\{F_x, F_y, F_z\} \times \{F_x, F_y, F_z\}$, excluding
the cases where $F = F'$, as those have been studied previously, and are not
under the alternative. Note that the ordering within the pair matters, because
the two graphs are not of the same order. We generate $1000$ graphs for each of
the possible combination of of $F$, $F'$, $c$, and $n$. We embed the graphs in
one dimension using ASE or CASE, use orthogonal alignment via the median trick,
and perform the nonparametric test. Previously, we were omitting the power
simulation that uses the invalid version of the test, since a test that doesn't
need to be valid can be arbitrarily powerful (simply, consider a test that
always rejects). However, we include the results here because they demonstrate
an important point regarding the using the uncorrected verions of the test.

The power of the test at significance level $\alpha=0.05$ in these six possible
settings is summarized in Figure \ref{fig:nonpar-power-rdpg}. There are several
important observations we can make. First, consider the right pannel which
summarizes the emperical power of the test that uses CASE. The power
monotonically increases as $n$ grows, hence the test for the latent distribution
that uses CASE is able to meaningfully distinguish between MMSBMs with different
amounts of mixing. Furthermore, observe that the settings of $F_x$ vs. $F_z$ and
$F_z$ vs. $F_x$, have more power than the all over ones. Thus, the power in the
setting when one mixture has a lot of mixing and the other has no mixing at all
is larger than the power in the setting of no versus some mixing, or in the
setting with some versus a lot of mixing, as one would expect.

Next, consider the left pannel, which presents the power using the invalid
version of the test that uses ASE. Observe that for some settings the power
decreases when using the corrected version of the test decreases, but increases
for others. In particular, it decreases the for settings of $F_y$ vs. $F_x$,
$F_z$ vs. $F_x$, and $F_z$ vs. $F_y$, and increases for the settings of $F_x$
vs. $F_y$, $F_x$ vs. $F_y$, and $F_y$ vs. $F_z$. To summarize, power decreases
in the cases when the smaller graph has more mixing than the larger graph, and
increases when the smaller graph has less mixing. Our conjecture is that the
increase of in power with the correction can happen if the distribution of the
latent positions of the smaller graph has smaller variance, as happens in the
cases where the smaller graph has more mixing. In over words, the difference in
variance due to the inherent differences in latent distributions is partially
compensated by the difference in variance due to estimation, which leads to less
powerful test if the correction is not used. Thus, using the uncorrected version
of the test can both lead to incorrect inference under the null hypothesis, and
a less sensitive inference under some alternatives.

\section{Real World Application}
\label{sec:real_data}
We demonstrate an application of this testing procedure to a real world dataset
of human connectomes. A connectome, also known as a brain graph
\citep{connectomics}, represents the brain as a network with neurons (or
collections thereof) as vertices, and synapses (or structural connections) as
edges. For this demonstration, the raw data is collected by diffusion magnetic
resonance imaging (dMRI), which can represent the structural connectivity within
the brain. \citep{real-data-example} This example is predominantly included as an
illustrative example of the applicability of the test to the setting and
consistency of its results with a natural intuition. It should not be treated
as an imaging study to draw conclusions about the dataset.

The macro-scale connectomes are estimated by NeuroData's MRI to graphs (NDMG)
pipeline \citep{real-data-pipeline}, which is designed to produce robust and
biologically plausible connectomes across studies, individuals, and scans. The
vertices of the graph represent regions of interest identified by spatial
proximity, and the edges of the graph represent the connection between regions
via tensor-based fiber streamlines. Specifically, there is an edge for a pair of
regions if and only if there is a streamline passing between them. For more
information on the procedure that generates the brain graphs, we refer the
readers to \cite{real-data-pipeline}. The data used in this study is the same
one used by \cite{real-data-example}.

Graphs in this dataset are weighted, directed, and have unknown dimension of the
latent distribution. Thus, modifications to the procedure described in
Subsection \ref{sec:generalizations} are required to obtain the ASE of the
graphs. The correction of ASE to CASE and the subsequent test are performed
without further modifications. In addition to that, we do employ the median
heuristic \citep{median-heuristic} in order to determmine the bandwidth of the
kernel. All of those modifications are implemented in {\bf graspologic}
\citep{graspy} Python package, and, in fact, used by default whenever one uses a
latent distribution test on a graphs that are directed, weighted, and/or have
unspecified latent dimension.

There are 57 subjects in this dataset, each of which has 2 different dMRI scans.
Furthermore, each of the scans was converted to a 'large' and a 'small' graph,
using the aforementioned pipeline. The number of vertices in the large graphs
varies between 730 and 1194, wheras the number of vertices in the small graphs
varies between 493 and 814. We will refer to these large and small graphs as
different scales. In this work all comparisons, whether within or between the
subjects, take two graphs of different scales, one small and one large.

\begin{figure}[t!]
  \begin{center}
    \includegraphics[width=0.95\textwidth]{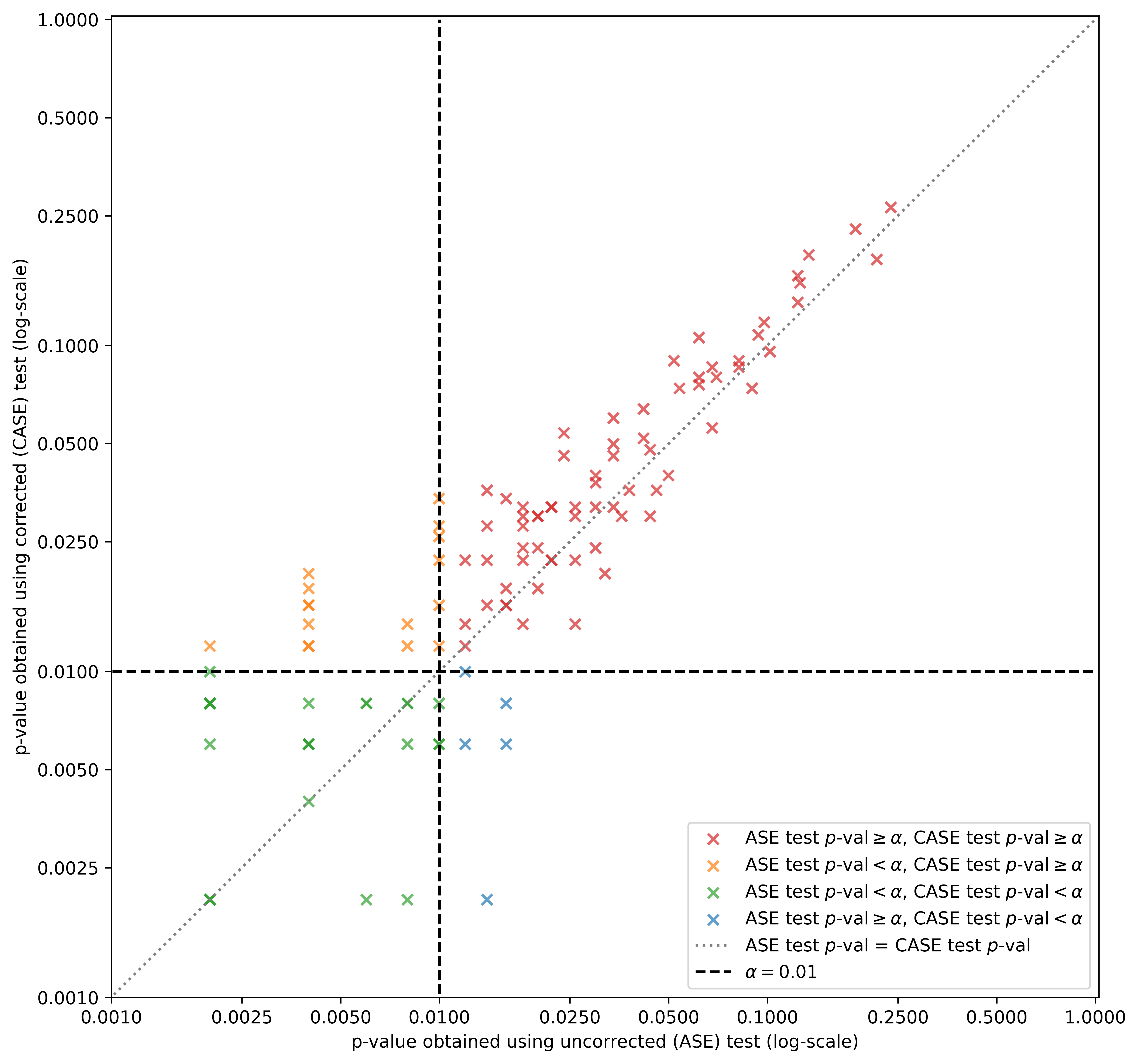}
  \end{center}
  \caption{Comparison of the difference in $p$-values obtained using CASE and
    ASE in the setting of brain graphs of the same subject and scan but
    different scales. Color coding according to decision at $\alpha=0.01$.
    Note that some datapoints might overlap exactly due to permutation test
    providing a $p$-value from a discrete set.}
  \label{fig:real_data_null}
\end{figure}

We first use both the test that uses ASE and the test that uses CASE to compare
the scans within the same subject, within the same scan, but between the two
scales. There are 114 total possible comparisons. Paired differences in
$p$-values obtained by the test that uses CASE and the test that uses ASE are
presented in Figure \ref{fig:real_data_null}. Using the one-sided Wilcoxon
Signed-Rank test \citep{wilcoxon} on those pairs of $p$-values, we obtain
$p$-value $<10^{-7}$, signifying that the corrected test rejects statistically
less often than the uncorrected test.

We can furthermore consider decision-theoretic consequences by setting
significance level $\alpha$ to two different commonly used values $0.01$ and
$0.05$. In case of $\alpha=0.01$, both tests reject the null in 24 case, neither
rejects in 69 cases, only ASE does in 16 cases, and only CASE does in just 5
cases (see color coding of Figure \ref{fig:real_data_null} but note that some
data points overlap). Using the two-sided Fisher's exact test, we obtain a
statistic of 20.7 with a $p$-value $<10^{-8}$. Alternatively, setting
$\alpha=0.05$, we obtain, a contigency table of: both: 89, none: 21, ASE only: 4,
CASE only: 0, leading to a two-sided Fisher's test statistic $p$-value
$<10^{-18}$. 

Thus, the test that uses CASE picks up the differences staistically less often
when using both raw $p$-values and when using binary decisions by comparing
$p$-values with significance levels $\alpha = \{0.01, 0.05\}$. In Section 4.3 we
demonstrated that using the uncorrected test can lead both to an invalid test
under the null, and to a less sensitive test under some alternatives. Thus using
a correction does not always imply having larger $p$-values. In this case,
however, it does, which aligns with our natural intuition that a correct test
should reject less often, as graphs obtained from the same scan but at different
scales should be somewhat similar to each other.

\begin{figure}[t]
  \begin{center}
    \includegraphics[width=\textwidth]{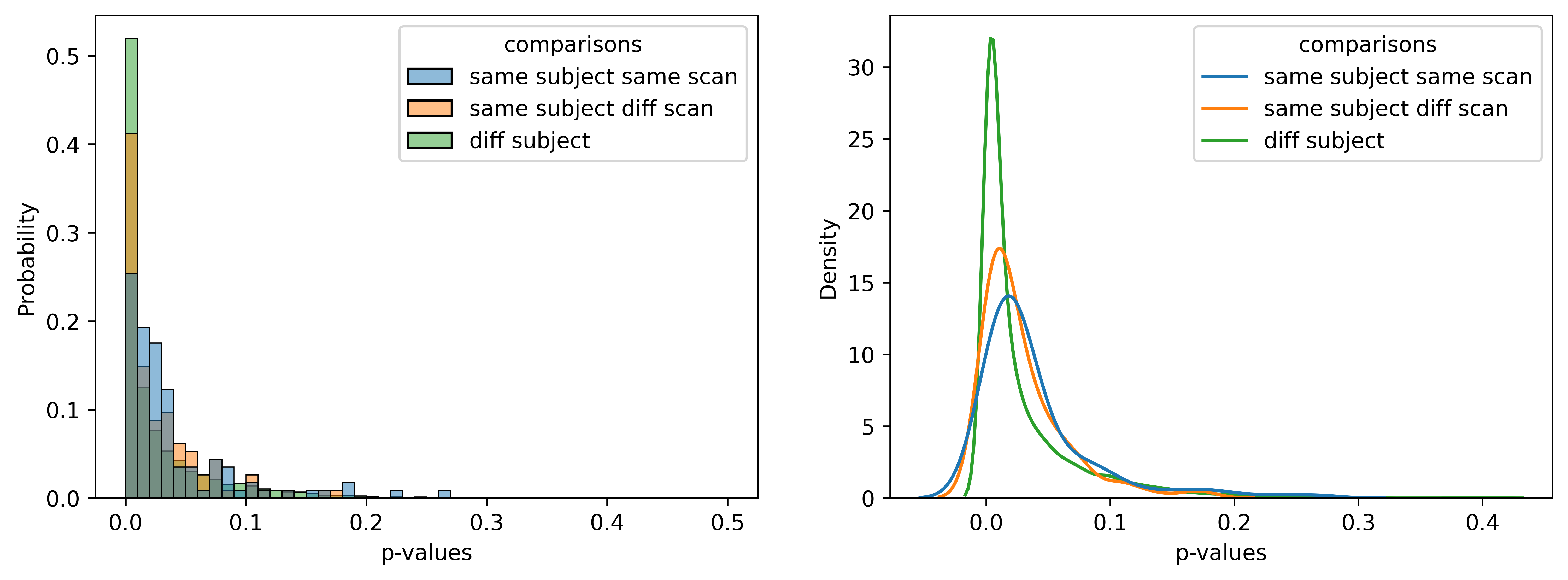}
  \end{center}
  \caption{Comparison of the difference in $p$-values obtaind using CASE for
    different settings of the brain graph data. Left panel: Histograms of
    $p$-values with a bin size of 0.01. Normalized to add to 1. Right panel:
    Kernel density estiates of the distribution of $p$-values. Normalized to
    integrate to 1.}
  \label{fig:real_data_alternative}
\end{figure}

Next, we use the corrected test to compare the graphs between the scans and
between subjects. Thre is a total of 114 possible comparisons in the setting of
different scans of the same subject, as there are two comparisons per subject
(larger scale scan 1 to smaller scale scan 2 and larger scale scan 2 to smaller
scale scan 1) and 57 subjects total. For the case of different subject, there
are $(57 \times 2) \times (56 \times 2) = 12768$ total comparions (57 subjects
each of which has 2 scans at larger scale compared to each of the 2 scans at
smaller scale of everyone but themselves).

We plot the histograms and the kernel density estimates of the distribution of
$p$-values, stratified by setting, in Figure \ref{fig:real_data_alternative}.
Using the one-sided ($>$) Mann-Whitney $U$ test \citep{mann-whitney} we obtain:
a $p$-value $0.020$ when comparing the distribution of $p$-values of the same
subject within the scan to the distribution of $p$-values for the same subject
between the scans, a $p$-value of $0.003$ when comparing the distribution of
$p$-values of the same subject between the scans to the distribution of
$p$-values between different subjects, and lastly, a $p$-value $<10^{-7}$ when
comparing the distribution of $p$-values for the same subject within the scan to
the distribution of $p$-values in the setting of different subjects.

To summarize, the $p$-values within the same subject same scan are smaller than
within the same subject but different scan, which are themselves smaller than
between different subjects. This aligns with the natural intuition that the test
should reject more often for different scans than for the same scans and more
often for the different subjects than for the same subjects.

\section{Discussion}
\label{sec:conclusion}
In this work we demonstrated that the latent distribution test proposed by
\cite{nonpar} degrades in validity as the numbers of vertices in two graphs
diverge from each other. This phenomenon does not contradict the results of the
original paper, as it occurs when test is used on two graphs of finite size.
Meanwhile, the scope of the original paper is limited to the asymptotic case.

We presented an intuitive example that demonstrates that the invalidity occurs
because a pair of adjacency spectral embeddings for the graphs with different
number of vertices falls under the alternative hypothesis of the subsequent
test. We also proposed a procedure to modify the embeddings in a way that makes
them exchangeable under the null hypothesis. This leads to a testing procedure
that is both valid and consistent, as has been demonstrated experimentally. The
code for the testing procedure that uses CASE is incorporated into {\bf GraSPy}
\cite{graspy} python package, alongside the original unmodified test. We
strongly recommend CASE, as opposed to ASE, for nonparametric two-sample graph
hypothesis testing when the graphs have differing numbers of vertices. However,
we note that this procedure is nondeterministic, as it requires sampling
additive noise.

Our work can be extended by developing limit theory for the corrected adjacency
spectral embeddings and the test statistcs that use them. It is also likely that
the approach of modifying the embeddings can be extended to tests that use
Laplacian spectral embedding (See \cite{rdpg-survey} for associated RDPG theory)
or models that are more general than RDPGs, such as Generalized Random Dot
Product Graphs \citep{grdpg} or other latent position models.

In general, two-sample latent distributon hypothesis testing is also closely
related to the problem of testing goodness-of-fit of the model \citep{nonpar}.
No such test, at least known to us, exists for random dot product graphs. We
hope that the work presented in this paper may facilitate this investigation.

\section{Declarations}
\subsection{Availability of data and materials}
\begin{itemize}
  \item The implementation of the latent distribution test used throughout the
    manuscript is now a part of {\bf graspologic} \citep{graspy} Python package,
    which is available at
    \href{https://github.com/microsoft/graspologic}{https://github.com/microsoft/graspologic}.
  \item The code used to run simulations, perform real data exeriments, and
    generate all figures in this manuscript is available at
    \href{https://github.com/alyakin314/correcting-nonpar}{https://github.com/alyakin314/correcting-nonpar}.
    The simulation results obtained by the authors are also provided in that
    repository.
  \item The dMRI dataset used and analysed in the Section \ref{sec:real_data}
    is available from the corresponding author on reasonable request.
\end{itemize}
\subsection{Competing interests}
The authors declare that they have no competing interests.
\subsection{Funding}
This was work was partially supported by the Defense Advanced Research Programs
Agency (DARPA) through the “Data-Driven Discovery of Models” (D3M) Program and
by funding from Microsoft Research.
\subsection{Authors' contributions}
\begin{itemize}
  \item AAA conceived the idea of the study, developed theory and methodology,
    created a software implementation, designed, programmed, and analyzed the
    synthetic and real data experiments, and drafted and edited the manuscript.
  \item JA developed theory and methodology, designed and analyzed the synthetic
    and real data experiments, and edited the manuscript.
  \item HSH developed theory and methodology, designed and analyzed the
    synthetic and real data experiments, and edited the manuscript.
  \item CEP conceived the idea of the study, developed theory and methodology,
    designed and analyzed the synthetic and real data experiments, and edited
    the manuscript.
\end{itemize}







\bibliographystyle{Chicago}

\bibliography{refs}

\begin{thebibliography}{}

\bibitem[\protect\citeauthoryear{Agterberg, Tang, and Priebe}{Agterberg
  et~al.}{2020}]{on-two-sources}
Agterberg, J., M.~Tang, and C.~E. Priebe (2020).
\newblock On two distinct sources of nonidentifiability in latent position
  random graph models.
\newblock arXiv:2003.14250.

\bibitem[\protect\citeauthoryear{Airoldi, Blei, Fienberg, and Xing}{Airoldi
  et~al.}{2008}]{mmsbm}
Airoldi, E.~M., D.~M. Blei, S.~E. Fienberg, and E.~P. Xing (2008).
\newblock Mixed membership stochastic blockmodels.
\newblock {\em Journal of Machine Learning Research\/}~{\em 9}, 1981–2014.

\bibitem[\protect\citeauthoryear{Arroyo, Athreya, Cape, Chen, Priebe, and
  Vogelstein}{Arroyo et~al.}{2021}]{graph-comparison-same-4}
Arroyo, J., A.~Athreya, J.~Cape, G.~Chen, C.~E. Priebe, and J.~T. Vogelstein
  (2021).
\newblock Inference for multiple heterogeneous networks with a common invariant
  subspace.

\bibitem[\protect\citeauthoryear{Asta and Shalizi}{Asta and
  Shalizi}{2015}]{graph-comparison-other-2}
Asta, D.~M. and C.~R. Shalizi (2015).
\newblock Geometric network comparisons.
\newblock In {\em Proceedings of the Thirty-First Conference on Uncertainty in
  Artificial Intelligence}, UAI'15, Arlington, Virginia, United States, pp.\
  102--110. AUAI Press.

\bibitem[\protect\citeauthoryear{Athreya, Fishkind, Tang, Priebe, Park,
  Vogelstein, Levin, Lyzinski, Qin, and Sussman}{Athreya
  et~al.}{2018}]{rdpg-survey}
Athreya, A., D.~E. Fishkind, M.~Tang, C.~E. Priebe, Y.~Park, J.~T. Vogelstein,
  K.~Levin, V.~Lyzinski, Y.~Qin, and D.~L. Sussman (2018).
\newblock Statistical inference on random dot product graphs: a survey.
\newblock {\em Journal of Machine Learning Research\/}~{\em 18\/}(226), 1--92.

\bibitem[\protect\citeauthoryear{Athreya, Priebe, Tang, Lyzinski, Marchette,
  and Sussman}{Athreya et~al.}{2016}]{avanti-clt}
Athreya, A., C.~E. Priebe, M.~Tang, V.~Lyzinski, D.~J. Marchette, and D.~L.
  Sussman (2016).
\newblock A limit theorem for scaled eigenvectors of random dot product graphs.
\newblock {\em Sankhya A\/}~{\em 78\/}(1), 1--18.

\bibitem[\protect\citeauthoryear{Bickel and Doksum}{Bickel and
  Doksum}{2006}]{Bickel+Doksum}
Bickel, P. and K.~Doksum (2006).
\newblock {\em Mathematical Statistics 2e}.
\newblock Pearson Education, Limited.

\bibitem[\protect\citeauthoryear{Bickel and Sarkar}{Bickel and
  Sarkar}{2016}]{graph-comparison-other-4}
Bickel, P.~J. and P.~Sarkar (2016).
\newblock Hypothesis testing for automated community detection in networks.
\newblock {\em Journal of the Royal Statistical Society Series B\/}~{\em
  78\/}(1), 253--273.

\bibitem[\protect\citeauthoryear{Chen and Lei}{Chen and
  Lei}{2018}]{graph-comparison-other-6}
Chen, K. and J.~Lei (2018).
\newblock Network cross-validation for determining the number of communities in
  network data.
\newblock {\em Journal of the American Statistical Association\/}~{\em
  113\/}(521), 241--251.

\bibitem[\protect\citeauthoryear{Chung, Bridgeford, Arroyo, Pedigo, Saad-Eldin,
  Gopalakrishnan, Xiang, Priebe, and Vogelstein}{Chung
  et~al.}{2021}]{connectomics}
Chung, J., E.~Bridgeford, J.~Arroyo, B.~D. Pedigo, A.~Saad-Eldin,
  V.~Gopalakrishnan, L.~Xiang, C.~E. Priebe, and J.~T. Vogelstein (2021).
\newblock Statistical connectomics.
\newblock {\em Annual Review of Statistics and Its Application\/}~{\em 8\/}(1),
  463--492.

\bibitem[\protect\citeauthoryear{Chung, Pedigo, Bridgeford, Varjavand, Helm,
  and Vogelstein}{Chung et~al.}{2019}]{graspy}
Chung, J., B.~D. Pedigo, E.~W. Bridgeford, B.~K. Varjavand, H.~S. Helm, and
  J.~T. Vogelstein (2019).
\newblock Graspy: Graph statistics in python.
\newblock {\em Journal of Machine Learning Research\/}~{\em 20\/}(158), 1--7.

\bibitem[\protect\citeauthoryear{Chung, Varjavand, Arroyo-Relión, Alyakin,
  Agterberg, Tang, Priebe, and Vogelstein}{Chung et~al.}{2022}]{varjavand}
Chung, J., B.~Varjavand, J.~Arroyo-Relión, A.~Alyakin, J.~Agterberg, M.~Tang,
  C.~E. Priebe, and J.~T. Vogelstein (2022).
\newblock Valid two-sample graph testing via optimal transport procrustes and
  multiscale graph correlation with applications in connectomics.
\newblock {\em Stat\/}~{\em 11\/}(1), e429.

\bibitem[\protect\citeauthoryear{de~Solla~Price}{de~Solla~Price}{1965}]{document-application-1}
de~Solla~Price, D.~J. (1965).
\newblock Networks of scientific papers.
\newblock {\em Science\/}~{\em 149\/}(3683), 510--515.

\bibitem[\protect\citeauthoryear{Erd\"{o}s and R\'{e}nyi}{Erd\"{o}s and
  R\'{e}nyi}{1960}]{er-graphs-3}
Erd\"{o}s, P. and A.~R\'{e}nyi (1960).
\newblock On the evolution of random graphs.
\newblock {\em Publications of the Mathematical Institute of the Hungarian
  Academy of Sciences\/}~{\em 5}, 17--61.

\bibitem[\protect\citeauthoryear{Escoufier}{Escoufier}{1973}]{rv-1}
Escoufier, Y. (1973).
\newblock Le traitement des variables vectorielles.
\newblock {\em Biometrics\/}~{\em 29\/}(4), 751--760.

\bibitem[\protect\citeauthoryear{Fan, Fan, Han, and Lv}{Fan
  et~al.}{2022}]{dcmmsbm-2}
Fan, J., Y.~Fan, X.~Han, and J.~Lv (2022).
\newblock Simple: Statistical inference on membership profiles in large
  networks.

\bibitem[\protect\citeauthoryear{Gangrade, Venkatesh, Nazer, and
  Saligrama}{Gangrade et~al.}{2019}]{graph-comparison-other-7}
Gangrade, A., P.~Venkatesh, B.~Nazer, and V.~Saligrama (2019).
\newblock Efficient near-optimal testing of community changes in balanced
  stochastic block models.
\newblock In {\em Advances in Neural Information Processing Systems 32}, pp.\
  10364--10375. Curran Associates, Inc.

\bibitem[\protect\citeauthoryear{Garreau, Jitkrittum, and Kanagawa}{Garreau
  et~al.}{2017}]{median-heuristic}
Garreau, D., W.~Jitkrittum, and M.~Kanagawa (2017).
\newblock Large sample analysis of the median heuristic.
\newblock arXiv:1707.07269.

\bibitem[\protect\citeauthoryear{Ghoshdastidar, Gutzeit, Carpentier, and von
  Luxburg}{Ghoshdastidar et~al.}{2017}]{graph-comparison-same-1}
Ghoshdastidar, D., M.~Gutzeit, A.~Carpentier, and U.~von Luxburg (2017, 07--10
  Jul).
\newblock Two-sample tests for large random graphs using network statistics.
\newblock In S.~Kale and O.~Shamir (Eds.), {\em Proceedings of the 2017
  Conference on Learning Theory}, Volume~65 of {\em Proceedings of Machine
  Learning Research}, Amsterdam, Netherlands, pp.\  954--977. PMLR.

\bibitem[\protect\citeauthoryear{Ghoshdastidar, Gutzeit, Carpentier, and von
  Luxburg}{Ghoshdastidar et~al.}{2020}]{graph-comparison-other-8}
Ghoshdastidar, D., M.~Gutzeit, A.~Carpentier, and U.~von Luxburg (2020).
\newblock {Two-sample hypothesis testing for inhomogeneous random graphs}.
\newblock {\em The Annals of Statistics\/}~{\em 48\/}(4), 2208 -- 2229.

\bibitem[\protect\citeauthoryear{Gilbert}{Gilbert}{1959}]{er-graphs-1-gilbert}
Gilbert, E.~N. (1959).
\newblock Random graphs.
\newblock {\em The Annals of Mathematical Statistics\/}~{\em 30\/}(4),
  1141--1144.

\bibitem[\protect\citeauthoryear{Gretton, Borgwardt, Rasch, Sch\"{o}lkopf, and
  Smola}{Gretton et~al.}{2012}]{gretton-survey}
Gretton, A., K.~M. Borgwardt, M.~J. Rasch, B.~Sch\"{o}lkopf, and A.~Smola
  (2012).
\newblock A kernel two-sample test.
\newblock {\em Journal of Machine Learning Research\/}~{\em 13}, 723--773.

\bibitem[\protect\citeauthoryear{Gretton, Fukumizu, Teo, Song, Sch\"{o}lkopf,
  and Smola}{Gretton et~al.}{2007}]{hsic}
Gretton, A., K.~Fukumizu, C.~H. Teo, L.~Song, B.~Sch\"{o}lkopf, and A.~J. Smola
  (2007).
\newblock A kernel statistical test of independence.
\newblock In {\em Proceedings of the 20th International Conference on Neural
  Information Processing Systems}, NIPS’07, Red Hook, NY, USA, pp.\
  585–592. Curran Associates Inc.

\bibitem[\protect\citeauthoryear{{Hardoon}, {Szedmak}, and
  {Shawe-Taylor}}{{Hardoon} et~al.}{2004}]{cca}
{Hardoon}, D.~R., S.~{Szedmak}, and J.~{Shawe-Taylor} (2004).
\newblock Canonical correlation analysis: An overview with application to
  learning methods.
\newblock {\em Neural Computation\/}~{\em 16\/}(12), 2639--2664.

\bibitem[\protect\citeauthoryear{Hoff, Raftery, and Handcock}{Hoff
  et~al.}{2002}]{latent-graphs-intro}
Hoff, P.~D., A.~E. Raftery, and M.~S. Handcock (2002).
\newblock Latent space approaches to social network analysis.
\newblock {\em Journal of the American Statistical Association\/}~{\em
  97\/}(460), 1090--1098.

\bibitem[\protect\citeauthoryear{Holland, Laskey, and Leinhardt}{Holland
  et~al.}{1983}]{sbm}
Holland, P.~W., K.~B. Laskey, and S.~Leinhardt (1983).
\newblock Stochastic blockmodels: First steps.
\newblock {\em Social Networks\/}~{\em 5\/}(2), 109 -- 137.

\bibitem[\protect\citeauthoryear{Jain, Duin, and Mao}{Jain
  et~al.}{2000}]{impossible}
Jain, A., R.~Duin, and J.~Mao (2000).
\newblock Statistical pattern recognition: a review.
\newblock {\em IEEE Transactions on Pattern Analysis and Machine
  Intelligence\/}~{\em 22\/}(1), 4--37.

\bibitem[\protect\citeauthoryear{Jin, Ke, and Luo}{Jin
  et~al.}{2017}]{dcmmsbm-1}
Jin, J., Z.~T. Ke, and S.~Luo (2017).
\newblock Estimating network memberships by simplex vertex hunting.
\newblock arXiv:1708.07852.

\bibitem[\protect\citeauthoryear{Karrer and Newman}{Karrer and
  Newman}{2011}]{dcsbm}
Karrer, B. and M.~E.~J. Newman (2011).
\newblock Stochastic blockmodels and community structure in networks.
\newblock {\em Physical Review E\/}~{\em 83\/}(1), 016107.

\bibitem[\protect\citeauthoryear{Kiar, Bridgeford, Roncal, for Reliability,
  (CoRR), Chandrashekhar, Mhembere, Ryman, Zuo, Margulies, Craddock, Priebe,
  Jung, Calhoun, Caffo, Burns, Milham, and Vogelstein}{Kiar
  et~al.}{2018}]{real-data-pipeline}
Kiar, G., E.~W. Bridgeford, W.~R.~G. Roncal, C.~for Reliability, R.~(CoRR),
  V.~Chandrashekhar, D.~Mhembere, S.~Ryman, X.-N. Zuo, D.~S. Margulies, R.~C.
  Craddock, C.~E. Priebe, R.~Jung, V.~D. Calhoun, B.~Caffo, R.~Burns, M.~P.
  Milham, and J.~T. Vogelstein (2018).
\newblock A high-throughput pipeline identifies robust connectomes but
  troublesome variability.
\newblock {\em bioRxiv\/}.

\bibitem[\protect\citeauthoryear{Lee, Shen, Priebe, and Vogelstein}{Lee
  et~al.}{2019}]{mgc-1}
Lee, Y., C.~Shen, C.~E. Priebe, and J.~T. Vogelstein (2019).
\newblock {Network dependence testing via diffusion maps and distance-based
  correlations}.
\newblock {\em Biometrika\/}~{\em 106\/}(4), 857--873.

\bibitem[\protect\citeauthoryear{Lei}{Lei}{2016}]{graph-comparison-other-3}
Lei, J. (2016).
\newblock A goodness-of-fit test for stochastic block models.
\newblock {\em Annals of Statistics\/}~{\em 44\/}(1), 401--424.

\bibitem[\protect\citeauthoryear{Lei}{Lei}{2018}]{graphons-2}
Lei, J. (2018).
\newblock Network representation using graph root distributions.
\newblock {\em Annals of Statistics (forthcoming)\/}.

\bibitem[\protect\citeauthoryear{{Levin}, {Athreya}, {Tang}, {Lyzinski}, and
  {Priebe}}{{Levin} et~al.}{2017}]{omni}
{Levin}, K., A.~{Athreya}, M.~{Tang}, V.~{Lyzinski}, and C.~E. {Priebe} (2017).
\newblock A central limit theorem for an omnibus embedding of multiple random
  dot product graphs.
\newblock In {\em 2017 IEEE International Conference on Data Mining Workshops
  (ICDMW)}, pp.\  964--967.

\bibitem[\protect\citeauthoryear{Levin and Levina}{Levin and
  Levina}{2019}]{graph-comparison-same-3}
Levin, K. and E.~Levina (2019).
\newblock Bootstrapping networks with latent space structure.
\newblock arXiv:1907.10821.

\bibitem[\protect\citeauthoryear{Li, Lei, Bhattacharyya, den Berge, Sarkar,
  Bickel, and Levina}{Li et~al.}{2020}]{hierarchical}
Li, T., L.~Lei, S.~Bhattacharyya, K.~V. den Berge, P.~Sarkar, P.~J. Bickel, and
  E.~Levina (2020).
\newblock Hierarchical community detection by recursive partitioning.
\newblock {\em Journal of the American Statistical Association\/}~{\em 0\/}(0),
  1--18.

\bibitem[\protect\citeauthoryear{Li and Li}{Li and
  Li}{2018}]{graph-comparison-same-2}
Li, Y. and H.~Li (2018).
\newblock Two-sample test of community memberships of weighted stochastic block
  models.
\newblock arXiv:1811.12593.

\bibitem[\protect\citeauthoryear{Lovász}{Lovász}{2012}]{graphons-1}
Lovász, L. (2012).
\newblock {\em Large Networks and Graph Limits.}, Volume~60 of {\em Colloquium
  Publications}.
\newblock American Mathematical Society.

\bibitem[\protect\citeauthoryear{Lyzinski, Sussman, Tang, Athreya, and
  Priebe}{Lyzinski et~al.}{2014}]{ase-consistency-3}
Lyzinski, V., D.~L. Sussman, M.~Tang, A.~Athreya, and C.~E. Priebe (2014).
\newblock Perfect clustering for stochastic blockmodel graphs via adjacency
  spectral embedding.
\newblock {\em Electronic Journal of Statistics\/}~{\em 8\/}(2), 2905--2922.

\bibitem[\protect\citeauthoryear{{Lyzinski}, {Tang}, {Athreya}, {Park}, and
  {Priebe}}{{Lyzinski} et~al.}{2017}]{hsbm}
{Lyzinski}, V., M.~{Tang}, A.~{Athreya}, Y.~{Park}, and C.~E. {Priebe} (2017).
\newblock Community detection and classification in hierarchical stochastic
  blockmodels.
\newblock {\em IEEE Transactions on Network Science and Engineering\/}~{\em
  4\/}(1), 13--26.

\bibitem[\protect\citeauthoryear{Mann and Whitney}{Mann and
  Whitney}{1947}]{mann-whitney}
Mann, H.~B. and D.~R. Whitney (1947).
\newblock On a test of whether one of two random variables is stochastically
  larger than the other.
\newblock {\em The Annals of Mathematical Statistics\/}~{\em 18\/}(1), 50--60.

\bibitem[\protect\citeauthoryear{Maugis, Olhede, Priebe, and Wolfe}{Maugis
  et~al.}{2020}]{graph-comparison-other-5}
Maugis, P.-A.~G., S.~C. Olhede, C.~E. Priebe, and P.~J. Wolfe (2020).
\newblock Testing for equivalence of network distribution using subgraph
  counts.
\newblock {\em Journal of Computational and Graphical Statistics\/}~{\em
  0\/}(0), 1--11.

\bibitem[\protect\citeauthoryear{Panda, Shen, Perry, Zorn, Lutz, Priebe, and
  Vogelstein}{Panda et~al.}{2021}]{exact-equivalence-2}
Panda, S., C.~Shen, R.~Perry, J.~Zorn, A.~Lutz, C.~E. Priebe, and J.~T.
  Vogelstein (2021).
\newblock Nonpar manova via independence testing.

\bibitem[\protect\citeauthoryear{Pearson}{Pearson}{1895}]{pearson}
Pearson, K. (1895).
\newblock Note on regression and inheritance in the case of two parents.
\newblock {\em Proceedings of the Royal Society of London\/}~{\em 58},
  240--242.

\bibitem[\protect\citeauthoryear{Priebe, Park, Tang, Athreya, Lyzinski,
  Vogelstein, Qin, Cocanougher, Eichler, Zlatic, and Cardona}{Priebe
  et~al.}{2017}]{neuroscience-application-1}
Priebe, C.~E., Y.~Park, M.~Tang, A.~Athreya, V.~Lyzinski, J.~T. Vogelstein,
  Y.~Qin, B.~Cocanougher, K.~Eichler, M.~Zlatic, and A.~Cardona (2017).
\newblock Semiparametric spectral modeling of the drosophila connectome.
\newblock arXiv:1705.03297.

\bibitem[\protect\citeauthoryear{Robert and Escoufier}{Robert and
  Escoufier}{1976}]{rv-2}
Robert, P. and Y.~Escoufier (1976).
\newblock A unifying tool for linear multivariate statistical methods: The rv-
  coefficient.
\newblock {\em Journal of the Royal Statistical Society. Series C (Applied
  Statistics)\/}~{\em 25\/}(3), 257--265.

\bibitem[\protect\citeauthoryear{Rubin-Delanchy}{Rubin-Delanchy}{2020}]{graphons-3}
Rubin-Delanchy, P. (2020).
\newblock Manifold structure in graph embeddings.
\newblock In H.~Larochelle, M.~Ranzato, R.~Hadsell, M.~F. Balcan, and H.~Lin
  (Eds.), {\em Advances in Neural Information Processing Systems}, Volume~33,
  pp.\  11687--11699. Curran Associates, Inc.

\bibitem[\protect\citeauthoryear{Rubin-Delanchy, Cape, Tang, and
  Priebe}{Rubin-Delanchy et~al.}{2022}]{grdpg}
Rubin-Delanchy, P., J.~Cape, M.~Tang, and C.~E. Priebe (2022).
\newblock A statistical interpretation of spectral embedding: The generalised
  random dot product graph.
\newblock {\em Journal of the Royal Statistical Society: Series B (Statistical
  Methodology)\/}~{\em 84\/}(4), 1446--1473.

\bibitem[\protect\citeauthoryear{Rubin-Delanchy, Priebe, and
  Tang}{Rubin-Delanchy et~al.}{2017}]{mmsbm-rdpg}
Rubin-Delanchy, P., C.~E. Priebe, and M.~Tang (2017).
\newblock Consistency of adjacency spectral embedding for the mixed membership
  stochastic blockmodel.
\newblock arXiv:1705.04518.

\bibitem[\protect\citeauthoryear{Rukhin and Priebe}{Rukhin and
  Priebe}{2011}]{graph-comparison-other-1}
Rukhin, A. and C.~E. Priebe (2011).
\newblock A comparative power analysis of the maximum degree and size
  invariants for random graph inference.
\newblock {\em Journal of Statistical Planning and Inference\/}~{\em 141\/}(2),
  1041 -- 1046.

\bibitem[\protect\citeauthoryear{Shen, Priebe, and Vogelstein}{Shen
  et~al.}{2020}]{mgc-2}
Shen, C., C.~E. Priebe, and J.~T. Vogelstein (2020).
\newblock From distance correlation to multiscale graph correlation.
\newblock {\em Journal of the American Statistical Association\/}~{\em
  115\/}(529), 280--291.

\bibitem[\protect\citeauthoryear{Shen and Vogelstein}{Shen and
  Vogelstein}{2021}]{exact-equivalence-1}
Shen, C. and J.~T. Vogelstein (2021).
\newblock The exact equivalence of distance and kernel methods in hypothesis
  testing.
\newblock {\em AStA Advances in Statistical Analysis\/}~{\em 105\/}(3),
  385--403.

\bibitem[\protect\citeauthoryear{Sussman, Tang, and Priebe}{Sussman
  et~al.}{2014}]{ase-consistency-2}
Sussman, D., M.~Tang, and C.~Priebe (2014).
\newblock Consistent latent position estimation and vertex classification for
  random dot product graphs.
\newblock {\em IEEE Transactions on Pattern Analysis and Machine
  Intelligence\/}~{\em 36}, 48--57.

\bibitem[\protect\citeauthoryear{Sussman, Tang, Fishkind, and Priebe}{Sussman
  et~al.}{2012}]{ase-consistency-1}
Sussman, D.~L., M.~Tang, D.~E. Fishkind, and C.~E. Priebe (2012).
\newblock A consistent adjacency spectral embedding for stochastic blockmodel
  graphs.
\newblock {\em Journal of the American Statistical Association\/}~{\em
  107\/}(499), 1119--1128.

\bibitem[\protect\citeauthoryear{Székely and Rizzo}{Székely and
  Rizzo}{}]{energy}
Székely, G.~J. and M.~L. Rizzo.
\newblock Energy statistics: A class of statistics based on distances.
\newblock {\em Journal of Statistical Planning and Inference\/}~{\em 143\/}(8),
  1249 -- 1272.

\bibitem[\protect\citeauthoryear{Székely and Rizzo}{Székely and
  Rizzo}{2014}]{dcorr-unbiased}
Székely, G.~J. and M.~L. Rizzo (2014).
\newblock Partial distance correlation with methods for dissimilarities.
\newblock {\em The Annals of Statistics\/}~{\em 42\/}(6), 2382--2412.

\bibitem[\protect\citeauthoryear{Székely, Rizzo, and Bakirov}{Székely
  et~al.}{2007}]{dcorr}
Székely, G.~J., M.~L. Rizzo, and N.~K. Bakirov (2007).
\newblock Measuring and testing dependence by correlation of distances.
\newblock {\em The Annals of Statistics\/}~{\em 35\/}(6), 2769--2794.

\bibitem[\protect\citeauthoryear{Tang, Athreya, Sussman, Lyzinski, Park, and
  Priebe}{Tang et~al.}{2017}]{semipar}
Tang, M., A.~Athreya, D.~L. Sussman, V.~Lyzinski, Y.~Park, and C.~E. Priebe
  (2017).
\newblock A semiparametric two-sample hypothesis testing problem for random
  graphs.
\newblock {\em Journal of Computational and Graphical Statistics\/}~{\em
  26\/}(2), 344--354.

\bibitem[\protect\citeauthoryear{Tang, Athreya, Sussman, Lyzinski, and
  Priebe}{Tang et~al.}{2017}]{nonpar}
Tang, M., A.~Athreya, D.~L. Sussman, V.~Lyzinski, and C.~E. Priebe (2017).
\newblock A nonparametric two-sample hypothesis testing problem for random
  graphs.
\newblock {\em Bernoulli\/}~{\em 23\/}(3), 1599--1630.

\bibitem[\protect\citeauthoryear{Tang, Cape, and Priebe}{Tang
  et~al.}{2022}]{minh-spectral}
Tang, M., J.~Cape, and C.~E. Priebe (2022).
\newblock {Asymptotically efficient estimators for stochastic blockmodels: The
  naive MLE, the rank-constrained MLE, and the spectral estimator}.
\newblock {\em Bernoulli\/}~{\em 28\/}(2), 1049 -- 1073.

\bibitem[\protect\citeauthoryear{Tang, Sussman, and Priebe}{Tang
  et~al.}{2013}]{rdpg-approximation}
Tang, M., D.~L. Sussman, and C.~E. Priebe (2013).
\newblock Universally consistent vertex classification for latent positions
  graphs.
\newblock {\em The Annals of Statistics\/}~{\em 41\/}(3), 1406--1430.

\bibitem[\protect\citeauthoryear{Wasserman and Faust}{Wasserman and
  Faust}{1994}]{social-application-1}
Wasserman, S. and K.~Faust (1994).
\newblock {\em Social network analysis: Methods and applications}, Volume~8.
\newblock Cambridge university press.

\bibitem[\protect\citeauthoryear{Wilcoxon}{Wilcoxon}{1945}]{wilcoxon}
Wilcoxon, F. (1945).
\newblock Individual comparisons by ranking methods.
\newblock {\em Biometrics Bulletin\/}~{\em 1\/}(6), 80--83.

\bibitem[\protect\citeauthoryear{Yang, Priebe, Park, and Marchette}{Yang
  et~al.}{2019}]{real-data-example}
Yang, C., C.~E. Priebe, Y.~Park, and D.~J. Marchette (2019).
\newblock Simultaneous dimensionality and complexity model selection for
  spectral graph clustering.
\newblock {\em Journal of Computational and Graphical Statistics\/}~{\em 30},
  422 -- 441.

\bibitem[\protect\citeauthoryear{Zhu and Ghodsi}{Zhu and
  Ghodsi}{2006}]{profile-likelihood}
Zhu, M. and A.~Ghodsi (2006, 02).
\newblock Automatic dimensionality selection from the scree plot via the use of
  profile likelihood.
\newblock {\em Computational Statistics and Data Analysis\/}~{\em 51},
  918--930.

\end{thebibliography}
\end{document}